\documentclass[preprint2]{aastex}

\input epsf.sty

\newcommand{\Msun}{M_{\odot}}

\def\gsim{\mathrel{\rlap{\lower 4pt \hbox{\hskip 1pt $\sim$}}\raise 1pt
\hbox {$>$}}}
\def\lsim{\mathrel{\rlap{\lower 4pt \hbox{\hskip 1pt $\sim$}}\raise 1pt
\hbox {$<$}}}

\begin{document}

\title{Three Dimensional Simulation of Gamma Ray Emission 
from Asymmetric Supernovae and Hypernovae} 

\author{
Keiichi Maeda
}
\affil{Department of Earth Science and Astronomy,
Graduate School of Arts and Science, University of Tokyo, Meguro-ku, Tokyo
153-8902, Japan
}

\begin{abstract}
Hard X- and $\gamma$-ray spectra and light curves resulting from 
radioactive decays are computed for aspherical (jet-like) and 
energetic supernova models (representing a prototypical hypernova 
SN 1998bw), using a 3D energy- and time-dependent Monte Carlo scheme. 
The emission is characterized by (1) early emergence of 
high energy emission, (2) large line-to-continuum ratio, 
and (3) large cut-off energy by photoelectric absorptions 
in hard X-ray energies. These three properties are not sensitively dependent 
on the observer's direction. On the other hand, 
fluxes and line profiles depend sensitively on the observer's direction, 
showing larger luminosity and larger degree of blueshift for an observer 
closer to the polar ($z$) direction. 
Strategies to derive the degree of asphericity and the observer's direction 
from (future) observations are suggested on the basis of these features, 
and an estimate on detectability of the high energy emission 
by the {\it INTEGRAL} and future observatories is presented. 
Also presented is examination on applicability of a gray 
effective $\gamma$-ray opacity for computing the energy deposition rate 
in the aspherical SN ejecta. The 3D detailed computations show that the 
effective $\gamma$-ray opacity $\kappa_{\gamma} \sim 0.025 - 0.027$ cm$^{2}$ g$^{-1}$ 
reproduces the detailed energy-dependent transport for both spherical and aspherical 
(jet-like) geometry. 
\end{abstract}


\section{INTRODUCTION}

Gamma rays emitted from radioactive isotopes, which are explosively synthesized at a 
supernova (SN) explosion (Truran, Arnett, \& Cameron 1967; Bodansky, Clayton, \& 
Fowler 1968; Woosley, Arnett, \& Clayton 1973), 
play an important and unique role in emission from a supernova, 
not only at $\gamma$-ray energies but also at the lower energies: from 
X-rays to even optical or near-infrared (NIR) band. 
Up to a few years after the explosion, the decay chain $^{56}$Ni $\to$ 
$^{56}$Co $\to$ $^{56}$Fe (Clayton, Colgate, \& Fishman 1969) 
dominates the high energy radiation input to a 
supernova, with minor contributions from other radioactivities such as $^{57}$Ni 
($\to$ $^{57}$Co $\to$ $^{57}$Fe: Clayton 1974). 
The decay produces $\gamma$-ray lines with average energy per decay 
$\sim 1.7$ MeV ($^{56}$Ni decay with the e-folding time 8.8 days) or 
$\sim 3.6$ MeV ($^{56}$Co decay with the e-folding time 113.7 days). 
These line $\gamma$-rays are degraded in their ways through the SN ejecta, predominantly 
by compton scatterings (see e.g., Cass\'e \& Lehoucq 1990 for a review). 
Non-thermal electrons produced at the scatterings and 
other processes (pair production and photoelectric absorption) 
heat the ejecta, yielding thermal emissions at optical and NIR wavelengths.

The $\gamma$-ray emissions from supernovae provide a unique tool to study the 
amount and distribution of predominant radioactive isotopes $^{56}$Co (therefore 
$^{56}$Ni produced at the explosion). 
Density structure in the supernova ejecta could also be inferred by modeling 
line profiles that are affected by compton scatterings. 
For example, unexpectedly early detections of the high energy emission 
from SN 1987A (Dotani et al. 1987; Sunyaev et al. 1987; Matz et al. 1988) 
revealed the important role of 
Rayleigh-Taylor instability and mixing in the SN ejecta 
(e.g., Chevalier 1976; Hachisu et al. 1990). 
The example highlights the very importance of modeling and analyzing 
$\gamma$-ray emission from supernovae. 
Except the very nearby SN 1987A, 
unfortunately there are to date only a few other examples of 
possible detection of $\gamma$-rays from the $^{56}$Ni decay chain 
from supernovae: 
One marginal detection (SN Ia 1991T: Lichti et a. 1994; 
Morris et al. 1997) and two upper limits (SNe Ia 1986G and 1998bu: 
Matz \& Share 1990; Leising et al. 1999). 
However, now that the {\it International Gamma-Ray Astrophysics Laboratory 
(INTEGRAL)} has been launched and some new gamma ray observatories are being planed 
(e.g., Takahashi, Kamae, \& Makishima 2001), 
it should be important to make prediction of $\gamma$-ray emissions (spectra and light curves), 
taking into account recent development of supernova researches, i.e., multi-dimensionality 
(see e.g., Maeda, Mazzali, \& Nomoto 2006b and references therein). 

Most of studies on $\gamma$-ray transport in SN ejecta 
have been restricted to one-dimensional, spherical models. 
Lately, a few 3D $\gamma$-ray transport computations for SNe have become available: 
H\"oflich (2002) developed a 3D $\gamma$-ray 
transport computational code and applied it to a 3D hydrodynamic model of 
a Type Ia supernova (Khoklov 2000). Hungerford, Fryer, \& Warren (2003) also 
developed a 3D $\gamma$-ray transfer code. They computed $\gamma$-ray spectra 
for 3D asymmetric (bipolar) type II supernova models, and later for 
3D single lobe explosion models (Hungerford, Fryer, \& Rockefeller 2005).  

Given higher density and smaller $^{56}$Ni mass for core-collapse supernovae than 
SNe Ia, there is no doubt that core-collapse supernovae are more difficult 
to detect in the high energy range (e.g., Timmes \& Woosley 1997). 
For example, H\"oflich, Wheeler, \& Khoklov (1998) predicted the 
detection limit of SNe Ia by {\it INTEGRAL} $\sim 10$ Mpc ($\sim 3$ per year), 
while Hungerford et al. (2003) estimated that for SNe II (more or less similar to SN 1987A) 
$\sim 650$ kpc ($\sim 1 - 2$ per 100 years). 
Accordingly, the number of theoretical prediction of 
hard X-ray and $\gamma$-ray emission from core-collapse supernovae is still 
small to date, except models for the very nearby SN 1987A (e.g., 
McCray, Shull, \& Sutherland 1987; Woosley et al. 1987; 
Shibazaki \& Ebisuzaki 1988; Kumagai et al. 1989). 
Despite this, in view of proposed large asymmetry in core-collapse supernovae 
(see e.g., Maeda \& Nomoto (2003b) and references therein) and 
its possible direct relation to high energy emissions, 
theoretical prediction of high energy emission for a 
variety of core-collapse supernova models is important in 
order to uncover the still-unclarified nature of the 
explosion and understand trends that may be possible to 
observe with current and future instruments.

In this respect, potentially interesting targets among core collapse 
supernovae are very energetic supernovae, often called "hypernovae" 
(Iwamoto et al. 1998). 
A prototypical hypernova is SN 1998bw discovered in association with 
a gamma ray burst GRB980425 (Galama et al. 1998). Its broad absorption features in 
optical spectra around maximum brightness and very bright peak magnitude 
led Iwamoto et al. (1998), assuming spherical symmetry, 
to conclude that the kinetic energy $E_{51} 
\equiv E_{\rm K}/10^{51}$ ergs $ \sim 30$, the ejecta mass $M_{\rm ej} 
\sim 10\Msun$, the main sequence mass $M_{\rm ms} \sim 40\Msun$, and 
the mass of newly synthesized radioactivity $^{56}$Ni 
$M$($^{56}$Ni) $\sim 0.6\Msun$ 
(see also Woosley, Eastman, \& Schmidt 1999). 
Following this and motivated by the deviation between the spherical hypernova model 
prediction and observations after $\sim 100$ days, 
Maeda et al. (2006ab) 
have presented comprehensive study comparing various 
observations of SN 1998bw with theoretical expectations 
from jet-like aspherical explosion models of Maeda et al. (2002),  
using multi-dimensional radiation transport calculations. 
They found that an aspherical model with $E_{51} \sim 20$ provides 
a good reproduction of optical emission from the explosion to $\gsim 1$ year 
consistently. 

This paper follows the analysis of optical emission from aspherical 
hypernovae applied to SN 1998bw by Maeda et al. (2006b). 
In this paper we present theoretical predictions of high 
energy emission from the same set of aspherical hypernova 
models as presented in Maeda et al. (2002, 2006ab).
Because the models are intrinsically aspherical, we make use of 
fully 3D hard X- and $\gamma$-ray transport computations. 
In \S 2, details of computational method are presented with a brief summary 
of the input models. In \S 3, overall synthetic spectra are presented. 
\S 4 focuses on line profiles. In \S 5, light curves of some lines 
are presented. 

In addition to modeling high energy emission from SN 1998bw-like hypernovae, 
also interesting is applicability of a gray effective absorptive opacity for 
$\gamma$-ray transport, which is often used as an approximation 
in computations of optical 
spectra and light curves of supernovae to save computational time 
(e.g., Sutherland \& Wheeler 1984). 
Although it has been examined in 1D spherically symmetric cases and concluded that 
the approximation is good if an appropriate value is used for the effective 
$\gamma$-ray opacity, 
it has not been examined yet if this is also the case for aspherical models. 
In \S 6, applicability of the gray absorptive $\gamma$-ray opacity for 
(jet-like) aspherical models 
is examined by comparing results with detailed $\gamma$-ray transport and 
with gray transport. 
Finally, in \S 7 conclusions and discussion, including an estimate of 
detectability of the high energy emission, 
are presented.

\section{METHOD AND MODELS}

\subsection{Method}

We have developed a fully 3D, energy-dependent, and time-dependent 
gamma ray transport computational code. It has been developed following 
the individual packet method using a Monte Carlo scheme 
as suggested by Lucy (2005). 
The code follows gamma ray transport in SN ejecta discretised 
in 3D Cartesian grids ($x_{i}, y_{j}, z_{k}$: linearly discretised) 
and in time steps ($t_{n}$: logarithmically discretised). 
For the ejecta dynamics, we assume homologous expansion, which should be 
a good approximation for SNe Ia/Ib/Ic. 
The density at time interval ($t_{n}, t_{n+1}$) 
is assumed homogeneous in each spatial zone with the value 
at time $t_{n+\frac{1}{2}} \equiv \sqrt{t_{n} t_{n+1}}$. 

The transport is solved in the SN rest frame. 
The expansion of the ejecta is taken into account as follows. 
First, cross sections for interactions with SN materials given in the 
comoving frame are transformed into the rest frame (Castor 1972). 
The fate of a photon is then determined in the rest frame. 
If a packet survives as high energy photons after the interaction, 
the new direction and the energy are given at the comoving 
frame, depending on the specific interaction taking place (see below). 
The direction and the energy are then converted to the SN rest frame 
(Castor 1972). 

Gamma ray lines from the decay chains $^{56}$Ni $\to$ $^{56}$Co $\to$ $^{56}$Fe 
and $^{57}$Ni $\to$ $^{57}$Co $\to$ $^{57}$Fe are included. 
The numbers of lines included in the computation 
are 6 ($^{56}$Ni decay), 24 ($^{56}$Co), 4 ($^{57}$Ni), and 3 ($^{57}$Co) 
(Lederer \& Shirley 1978; Ambwani \& Sutherland 1988). 
In the present study 
we focus on the $^{56}$Ni $\to$ $^{56}$Co $\to$ $^{56}$Fe chain, 
which includes 812 KeV ($^{56}$Ni), 847 keV ($^{56}$Co), and 
1238 keV ($^{56}$Co) lines. 

For the interactions of gamma rays 
with SN materials, we consider pair production, compton scattering, 
and photoelectric absorption. 
For pair production cross sections, a simplified fitting formula to Hubbell (1969) 
is adopted from Ambwani \& Sutherland (1988). 
For compton scattering, the Klein-Nishina cross section is used. 
For photoelectric absorption, cross sections compiled by H\"oflich, Khoklov, \& 
M\"uller (1992) from Veigele (1973) are used. 

If an interaction takes place, the fate of the gamma ray packet is chosen 
randomly in proportion to the cross section of each possible interaction. 
If the pair production is the fate, 
an electron and a positron are created. 
For the electron path, the electron deposits the entire energy to the ejecta. 
For the positron path, the positron annihilates with 
an ambient electron, producing two $\gamma$-rays (assuming no positronium formation). 
The positron kinetic energy is deposited to the thermal pool. 
These processes are assumed to take place {\it is situ}. 
According to this prescription, a packet either becomes 
the 511 keV $\gamma$-ray lines or is absorbed. 
The fate is selected randomly in proportion to the branching probability 
depending on the initial photon energy before the pair production. 
If the compton scattering is the fate, the polar angle of the scattering relative to 
the incoming photon direction in the comoving frame is randomly sampled 
according to the Klein Nishina distribution using a standard Monte Carlo rejection 
technique (the Kahn's method). The azimuthal angle is randomly selected in the comoving frame. 
The packet is now either a gamma-ray packet or a non-thermal electron packet (assumed to 
deposit the entire energy to the SN ejecta immediately), determined 
randomly according to the energy 
flowing into each branch. 
If the photoelectric absorption is the fate, the entire energy of the packet is 
absorbed. Possible X-ray fluorescence photons are not followed in the present simulations, 
but these photons are below the low energy cut-off, making negligible effects 
on the energy range examined in this paper. 

For a time advancing scheme, the code has two options: 
The one is the exact time advancing, taking into account the time interval 
at each photon flight. The other is no time advancing, assuming 
the flight time through the ejecta is negligible. 
In the present work, we use the exact one for computations of $\gamma$-ray 
light curves (and for computations of optical emission in 
Maeda et al. 2006b), 
while use the no-time advancing for computations of spectra. 
This is in order to optimize the number of photons escaping at 
a given time interval in the spectral computations, 
since the computation of spectra with fine energy bins already 
needs a large number of packets and large computational time. 
For $\gamma$-rays, the negligible time delay approximation is a good one. 
We have confirmed this for the present models, 
by performing the fully time-dependent computation 
but with the number of packets entering each time interval smaller 
than used in the standard no-time advancing spectrum computations.

In view of the recent investigation by Milne et al. (2004) that 
not all the published 1D gamma ray transport codes give mutually 
consistent results, we test the capability of our new code by computing 
gamma ray spectra based on the (spherical) SN Ia model W7 
(Nomoto, Thielemann, \& Yokoi 1984), 
for which many previous studies are available for comparison. 
We have compared our synthetic gamma ray spectra at 25 and 50 days with Figure 5 of 
Milne et al. (2004). We found an excellent agreement between our results 
and the spectra resulting from majority of previous codes, 
e.g., of Hungerford et al. (2003). 

In this study, the ejecta are mapped onto $60^{3}$ Cartesian grids. 
For spectrum synthesis, $1.5 \times 10^{8}$ photon packets with equal 
initial energy content are used. 
In escaping the eject, these packets are binned into 10 angular zones 
with equal solid angle from $\theta = 0^{\rm o}$ to $180^{\rm o}$ 
(here $\theta$ is the polar angle from the $z$-axis) and into 
3000 energy bins up to 3 MeV with equal energy interval 1 keV. 
For $\gamma$-ray light curve computations, $10^{8}$ photon packets 
are used. In escaping the ejecta, the packets 
are binned into 36 time step logarithmically spaced from day 5 to 
day 300, as well as into the angular and energy bins.  

\subsection{Models}

Input models for the $\gamma$-ray transport computations are 
taken from the aspherical model A and the spherical model F from 
Maeda et al. (2006b). 
Model A is a result of a jet-like explosion with the initial 
energy input at the collapsing core injected more in the $z$-axis 
than in the $r$-axis (see Maeda et al. 2002 for details).  
In Maeda et al. (2006ab), we examined optical light curves 
($\lsim 500$ days)
as well as expected optical spectral characteristics in both 
early ($\lsim 100$ days) and late ($\gsim 100$ days) phases. 
The structure of Model A at homologous expansion phases 
is shown in Figure 1.

In the present work we examine the following models: 
($M_{\rm ej}/\Msun$, $E_{51}$, $M$($^{56}$Ni)/$\Msun$) 
$ = (10.4, 10, 0.31)$ and $(10.4, 20, 0.39)$ for Model A, and 
$(10.4, 10, 0.28)$ and $(10.4, 50, 0.40)$ for Model F 
(hereafter $M_{\rm ej}$ is the ejecta mass, $E_{51}$ is the kinetic 
energy of the expansion in $10^{51}$ ergs, and $M$($^{56}$Ni) is 
the mass of $^{56}$Ni synthesized at the explosion). 
Maeda et al. (2006b) concluded that optical properties of 
SN 1998bw are explained consistently by Model A with the energy 
$E_{51} \sim 20$ and with $M$($^{56}$Ni) $\sim 0.4\Msun$, 
so that we regard Model A representing a prototypical hypernova. 

As seen in Figure 1, Model A is characterized by concentration of 
$^{56}$Ni distribution along the $z$-axis, which is a consequence of 
the explosive nucleosynthesis in jet-like aspherical explosions 
(Nagataki 2000; Maeda et al. 2002; Maeda \& Nomoto 2003b). 
Because $^{56}$Ni is the source of $\gamma$-rays,  
the distribution will affect the $\gamma$-ray transport and 
resulting hard X- and $\gamma$-ray emissions sensitively. 
Figure 2 shows the distribution of $^{56}$Ni 
along the line of sight for our models. 
The amount of $^{56}$Ni is integrated in the plane 
perpendicular to the line of sight within the constant 
line-of-sight velocity interval. 
Such that, Figure 2 shows a profile of $\gamma$-ray lines 
from the decay of $^{56}$Ni or $^{56}$Co in optically thin limit. 

Figure 2 shows that Model A yields the $^{56}$Ni at high velocities 
if it is viewed from the polar ($z$) direction. On the other hand, if 
it is viewed from the equatorial ($r$) direction, 
it yields the distribution sharply peaked at zero and low velocities. 
Note that Model A with $E_{51} = 20$ shows $^{56}$Ni 
at velocities higher than in Model F with $E_{51} = 50$.

\section{HARD X- AND $\gamma$-RAY SPECTRA}

Figures 3 and 4 show synthetic $\gamma$- and hard X-ray spectra 
at 25 days after the explosion. 
Figures 5 and 6 show the ones at 50 days after the explosion. 
Thanks to the large $E_{\rm K}/M_{\rm ej}$ as compared to (normal) 
SNe II and to the absence of a massive hydrogen layer, 
the emergence of the high energy emission is much 
earlier for the present models (in the order of a month) than SNe II 
(in the order of a year). 

An aspherical model yields the emergence of the 
high energy emission earlier than a spherical model with the same energy:  
At 25 days, Model A with $E_{51} = 10$ already shows high energy appearance 
although Model F with $E_{51} = 10$ does not show up. 
Indeed, the former, despite the energy $E_{51} = 10$, gives 
the date of emergence comparable to Model F with $E_{51} = 50$. 
The time scale of the $\gamma$-ray emission will be further 
discussed in \S 5. 

Model A is characterized by the large line-to-continuum ratio, especially at 
early on (Figure 3). Because the continuum is formed by 
degradation of line photons by compton scatterings, the large line-to-continuum 
ratio is realized for the plasma with low optical depth for compton scatterings. 
Indeed, $\gamma$-ray transport computations typically predict larger the ratio 
at more advanced epochs (see e.g., Sumiyenov et al. 1990). It is also 
seen by comparing Model F with different energies ($E_{51} = 10, 50$) at day 50 
(Figure 5): The larger energy leads to smaller optical depth, and therefore to 
the larger line-to-continuum ratio.  
At 25 days, Model A with either $E_{51} = 20$ or $10$ yields the ratio larger 
than Model F with $E_{51} = 50$, which is attributed to the existence of 
high velocity $^{56}$Ni at low optical depth in Model A (Figures 1 \& 2). 
At 50 days, the effect becomes less significant (i.e., the ratio becomes 
comparable for Model A ($E_{51} = 10$ or $20$) and for Model F with $E_{51} = 50$), 
but still visible as compared with Model F with $E_{51} = 10$ (Figure 5).  

Another feature is seen in hard X-ray spectra. 
Model A has the hard X-ray cut-off at energy higher than Model F (Figures 4 and 6). 
The cut-off is formed by photoelectric absorption, 
which is dependent on metal content (e.g., Grebenev \& Sunyaev 1987).
At 25 days, only the emission near the surface (along the $z$-axis for Model A) 
can escape out of the ejecta. The cut-off is therefore determined by metal 
content near the surface. In Model F, the surface layer is dominated by intermediate 
mass elements (i.e., a CO core of the progenitor star). On the other hand, 
in Model A the emitting region contains a large fraction of Fe-peak elements, 
most noticeably $^{56}$Ni (or Co, Fe). Therefore, the photoelectric cut-off 
should be at the energy higher in Model A than Model F. 
At more advanced epochs, an observer looks into deeper regions. 
Then the contribution from the deeper region becomes bigger and bigger, 
yielding increase of the cut-off energy in Model F since $^{56}$Ni is 
centrally concentrated (Figure 2). This effect is also seen in Model A, but to 
the smaller extent. Although the mass of $^{56}$Ni within a given 
velocity interval increases toward the center (except the inner most region where 
the $^{56}$Ni fraction is very small) also in Model A along the $z$-axis, 
the increase is less dramatic than model F (e.g., compare the masses of 
$^{56}$Ni at 20,000 and 10,000 
km s$^{-1}$ for Model A in Figure 2). Therefore, the metal content of 
the emitting region does not temporally evolve very much in Model A. 

In the above discussion on the photoelectric absorption, one would expect 
to use the $^{56}$Ni distribution along the "$r$-axis", not the $z$-axis, for the observer 
at the $r$-axis. However, it is not the case. 
As discussed by Hungerford et al. (2002, 2005), even for the observer at the $r$-axis, 
$^{56}$Ni contributing the most of the emission is that at the $z$-axis, i.e., 
an observer at the $r$-axis looks at the emitting polar $^{56}$Ni blobs sideways 
(see also Maeda et al. 2006b). 
To clarify this, in Figure 7 we show last scattering points of 
hard X- and $\gamma$-rays. Also shown are contours of the optical depth 
for observers at $+z$- and $+r$-directions. The observer at the $z$-axis 
looks at the emitting blob moving toward the observer, while the observer on the 
$r$-plane looks at a pair of the emitting blobs sideways. 

This is further discussed in \S 4 and 5, but here we point out one difference 
related to effects of the viewing angle in our model from Hungerford et al. (2002). 
Their models do not yield large difference in the absolute flux, 
while our models do show boost of high energy luminosity toward the $z$-axis. 
A similar behavior is also seen in optical emission in our models (Maeda et al. 2006b). 
The mechanism of the boost should be the same as the one for optical photons. 
Initially at high density only the sector-shaped region along the $z$-axis can yield escaping 
photons (Fig. 7). The cross sectional area of this photosphere is larger toward the $z$-axis 
than the $r$-axis, making the luminosity boost toward the $z$-axis. 
As time goes by, the ejecta optical depth decreases, and therefore 
the photosphere moves to cover the equatorial region. 
The difference becomes less and less significant, as seen by comparing Figures 
3 and 5. 
The different behavior is probably due to the different degree of the penetration of $^{56}$Ni 
into the outer layers. Hungerford et al. (2002, 2005) 
presented SN II models with a massive hydrogen envelope, 
which does not exist in our model SNe Ib/c. 
The existence of the hydrogen envelope generally yields less aspherical distribution 
of $^{56}$Ni and the emitting region in SNe II than SNe Ib/c (e.g., Wang et al. 2001). 
In any case, as Hungerford et al. (2005) suggested, the high energy emission 
can very sensitively depending on the type and degree of asymmetry, 
therefore transport calculations are very important.

\section{LINE PROFILES}

We now turn our attention to line profiles. 
Figure 8 shows line profiles of the $^{56}$Ni 812 keV and $^{56}$Co 
847 keV lines at 25 days. Figure 9 shows the $^{56}$Co 1238 keV line. 
Figures 10 and 11 show the same lines at 50 days. 
At these early epochs, the ejecta are not thin to high energy photons 
(Figure 7), 
and therefore the line profiles are different from the ones expected 
in optically thin limit (Figure 2). 

At day 25, an observer only sees the region near the surface (Figure 7). 
For Model A, the region is further concentrated along the $z$-axis 
(the top of the "$^{56}$Ni bubble" in Figure 1). 
Figures 8 and 9 show that the line profiles are very asymmetric, 
showing only the emission at the blue coming from the 
region moving toward an observer, {\it except} an observer 
at the $r$-axis. 
For an observer at the $r$-axis, even the emission at the rest wavelength 
is seen, which is especially evident for the more energetic, therefore less 
optically thick, model (Model A with $E_{51} = 20$). 
The emitting region is now at the outer edge(s) of the $^{56}$Ni distribution, 
and the density is relatively small there (Figure 1). The line-of-sight, connecting 
the emitting region(s) and an observer at the $r$-direction passes only 
through the low density region, including the region moving away from the observer. 
On the other hand, the line-of-sight for an observer at the $z$-axis 
passes through the dense, central region. 
For example, this behavior is seen in Figure 7 by comparing the regions, 
having the optical depth $\tau_{\gamma}$ 
less than 10 at day 25, for the observers at $z$- and $r$-axes. 
For the observer at $+z$ direction, the region contains only the emitting blob 
moving toward the observer. For the observer at the $+r$ direction, on the other hand, 
a pair of the emitting blobs along the $z$-axis, seen from the side by the observer, 
are included in the regions with $\tau_{\gamma} < 10$. 

At 50 days, the line becomes broad redward as a photon even from the far side 
escape out of the ejecta more easily because of decreasing density. 
Still, only the blue part is seen for an observer at the $z$-direction 
(Figures 10 \& 11). It is in contrast to Model F with $E_{51} = 50$, 
which now show the emission at the rest wave length. 
It is interesting to see that the line profiles of Model A (either $E_{51} = 10$ or $20$) 
viewed at the $z$-direction resemble to the $^{56}$Ni distribution 
in the hemisphere moving toward the observer (Figure 2). 
This suggests that the photons from the far hemisphere are almost totally 
blocked by the high density 
central region, while the $^{56}$Ni-rich region itself is nearly optically thin. 
It is indeed the case as seen in Figure 7. 
For Model A viewed at the $r$-direction and for Model F, the line profile is explained by 
continuously decreasing escape probability. 
These arguments imply that, assuming that we have a temporal series of $\gamma$-ray observations, 
the line profiles for Model A viewed at the $z$-direction should show the time interval 
within which the line profiles are almost fixed, while Model A viewed 
at the $r$-direction and Model F  
should show the continuous changes in the line profile until the entire 
ejecta become optically thin. 
This could be an interesting observational target, since the 
temporally "fixed" line profile, if observed,  
suggests that the viewing angle is close to the pole. In this case, the line shape 
directly traces the distribution of $^{56}$Ni. 
In addition, line profiles, sensitively dependent on the viewing angle as well as 
the degree of asphericity, could be used as a tracer of the distribution of $^{56}$Ni and 
density once $\gamma$-ray observations at $\sim 1$ MeV with sufficient sensitivity 
become possible. The detectability will be discussed in \S 7.

\section{LIGHT CURVES}

Figures 12 shows light curves of the $^{56}$Ni 812, $^{56}$Co 847, 
and $^{56}$Co 1238 keV lines at the distance 10 Mpc. 
For the 1238 keV line, the flux is computed by integrating 
a spectrum in 1150 -- 1340 keV range. For the 812 and 847 keV lines, 
the sum of the fluxes is computed first by integrating a spectrum 
in the 800 -- 920 keV range, then the individual contribution is computed 
with the decay probability at the corresponding epochs 
assuming that the escape fractions for these two lines are equal. 
The assumption of the equal escape fractions should be a good approximation, 
since these two lines are close in the energy, yielding almost identical 
compton cross sections (see also Milne et al. 2004). 

Model A shows the emergence of the $\gamma$-rays earlier, 
even though the energy is smaller, than Model F. 
This is due to the large amount of $^{56}$Ni at low density 
and high velocity regions. Later on, the $\gamma$-rays 
become stronger for Model F with $E_{51} = 50$ than for Model A 
($E_{51} = 10, 20$). Three effects are responsible:  
(1) The contribution of the 
dense central region stopping the photons becomes large. 
(2) The overall optical depth $\tau \propto 
M^2/E$ is larger for Model A than Model F. 
(3) Model F with $E_{51} = 50$ has a bit large $M$($^{56}$Ni) 
as compared to Model A. 
Because of this temporal behavior, the enhancement of $\gamma$-ray flux 
is more noticeable for the $^{56}$Ni line (with the e-folding time 8.8 days) 
than the $^{56}$Co lines (with the e-folding time 113.7days). 

As discussed in \S 3, the effect of the viewing angle is seen at 
early epochs. Model A emits more toward the $z$-direction than the $r$-direction. 
This behavior is qualitatively similar to that in optical emission (Maeda et al. 2006b). 
The effect virtually vanishes around $\sim 100$ days for Model A with $E_{51} = 20$, 
which is consistent 
with a rough estimate that the ejecta become optically thin to 
compton scatterings at $\sim \sqrt{\tau_{0}} \sim (1250 \times 
(M_{\rm ej}/\Msun)^2/E_{51})^{1/2} \sim 80$ days (See e.g., Maeda et al. 2003a: 
see also Figure 7), 
where $\tau_{0}$ is the optical depth to compton scatterings 
at day 1. In the above estimate, the cross section is assumed to be $1/6$ of the 
Thomson scattering cross section, and composition $Y_e = 0.5$.

\section{COMPARISON BETWEEN DETAILD AND GRAY TRASNPORT}

In previous sections, we presented model predictions for  
high energy spectra and light curves. 
However, $\gamma$-ray transport in the SN ejecta has 
another role which is as important as the high energy emission itself: 
the $\gamma$-rays give heating of the ejecta, i.e., 
non-thermal electrons scattered off from ions by compton scatterings 
rapidly pass their energies to thermal particles through ionization, 
excitation, and scatterings with thermal electrons. 
The thermal energy is then converted to optical photons. 
In this way, the energy lost from the high energy photons 
determines the optical luminosity from a supernova. 
In this section, we examine how much energy is stored 
in the ejecta. 

In Maeda et al. (2006b), we made use of the detailed 3D high energy photon 
transport to compute optical light curves and nebular spectra 
of Models A and F (and other models). 
On the other hand, a gray effective absorptive assumption for $\gamma$-ray transport 
has been frequently used for computations of $\gamma$-ray deposition and 
optical light curves. 
Sometimes it has been used, although in spherically symmetric models, 
to compute optical light curves for (hypothetical) non-spherically 
symmetric supernovae 
(e.g., Tominaga et al. 2005; Folatelli et al. 2006). 
It is therefore important to examine applicability of the assumption for 
aspherical supernovae. 

Figure 13 shows synthetic optical light curves for Models A and F. 
Results with the detailed 3D transport and with the simplified 
gray absorptive $\gamma$-ray opacity (with various effective opacity $\kappa_{\gamma}$) 
are compared. 
Figure 13 shows that the gray absorptive approximation is actually good 
for computations of $\gamma$-ray deposition, thus for 
computations of optical light curves, 
if an appropriate value is used for the effective opacity. 
With the value $\kappa_{\gamma} = 0.027$ cm$^{2}$ g$^{-1}$, we obtain 
optical light curves for both the spherical model F and 
the aspherical model A almost identical to those obtained with 
the detailed $\gamma$-ray transport computations. 

At late epochs $\gsim 100 - 200$ days (depending on models) 
the ejecta become optically thin to $\gamma$-rays. 
At these epochs, $\gamma$-rays suffer 
at most only one compton scattering before escaping the ejecta. 
In this idealized situation, the "effective" $\gamma$-ray opacity 
can be computed by taking an appropriate average of 
the Klein-Nishina cross section for various scattering angles 
and $\gamma$-ray line energies. 
In this way, Sutherland \& Wheeler (1984) yielded $\kappa_{\gamma} = 
0.022$ cm$^{2}$ g$^{-1}$ assuming the typical line energy $\sim 2$ MeV. 
This value is dependent on the line list. 
We have also performed the estimate of the effective opacity in the optically thin limit  
under the condition that the averaged absorbed energy per photon flight is equal 
for the detailed case and for the effective absorption case, 
and found that for the $^{56}$Co lines, $\kappa_{\gamma} = 5.08$ (Z/A) cm$^{2}$ g$^{-1}$. 
This yields $\kappa_{\gamma} \sim 0.025$ cm$^{2}$ g$^{-1}$ for the SNe Ib/c composition. 
Indeed, Figure 13 implies that this value is probably even better than 
$0.027$ cm$^{2}$ g$^{-1}$ in the late phase, 
although the difference is small. 
We emphasize that in this situation, the effective opacity is independent 
from geometry of the ejecta. It is confirmed, although not generally, 
by the fact that the $\kappa_{\gamma} = 0.027$ cm$^{2}$ g$^{-1}$ 
reproduces the light curve obtained by the detailed computations for both Models A and F. 

Somewhat surprising is that (1) the same value $\kappa_{\gamma} = 0.027$ 
cm$^{2}$ g$^{-1}$ reproduces the detailed transport computations rather well 
at early epochs, 
and (2) this applies not only to the spherical model, but also to the aspherical model. 
Because multiple scatterings take place, now the effective $\gamma$-ray 
opacity can be geometry- and time-dependent, and can be different from the value 
at the late phases. 
For example,  
Sutherland \& Wheeler suggested $\kappa_{\gamma} = 0.03$ cm$^{2}$ g$^{-1}$ 
taking into account multiple scatterings. 
Frannson (1990) gave the value 
$\kappa_{\gamma} = 0.03$ cm$^{2}$ g$^{-1}$ with $Y_e = 0.5$ 
from a series of Monte Carlo simulations. 
Colgate, Petschek, \& Kriese (1980) gave the value 
$\kappa_{\gamma} = 0.028$ cm$^{2}$ g$^{-1}$. 
These authors used different ejecta models. 
The small differences among these studies 
(although there are some exceptions, e.g., 0.07 cm$^{2}$ g$^{-1}$ 
by Woosley, Taam, \& Weaver (1986)) suggest that 
the effects of density and $^{56}$Ni distribution on the total deposition rate is 
small (in spherically symmetric case). 
The present study suggests that this is also the case even in asymmetric cases.

\section{CONCLUSIONS AND DISCUSSION} 

In this paper, we first reported the development of a new code 
for 3D computations of hard X-ray and $\gamma$-ray emission 
resulting from radioactive decays in supernovae. 
The code is useful for many problems. 
Among them is applicability of a gray absorptive approximation in supernova 
ejecta without spherically symmetry. We have shown, although only for some specific models, 
this is actually a good approximation. 
This approximation has been extensively used even in 
analyzing supernovae with hypothetical asymmetry (e.g., 
Tominaga et al. 2005; Folatelli et al. 2006), 
and we have confirmed the applicability using the 3D computations. 

With the code, we have presented predictions of 
hard X-ray and $\gamma$-ray emission in aspherical hypernova models. 
The models were verified by fitting optical 
properties of the prototypical hypernova SN 1998bw (Maeda et al. 2006ab), 
therefore we regard the model prediction being based on "realistic" hypernovae. 

Irrespective of the observer's direction, 
the aspherical models yield (1) early emergence of the high energy emission, 
and therefore the large peak flux, 
(2) large line-to-continuum ratio, and (3) large cut-off energy in the hard 
X-ray band, as compared to spherical models. 
These are qualitatively similar to what are expected from extensive mixing 
of $^{56}$Ni (see e.g., Cass\"e \& Lehoucq 1990) 
by e.g., Rayleigh-Taylor instability (e.g., Hachisu et al. 1990). 
However, degree of the $^{56}$Ni penetration is more extreme, and therefore 
the effects of $^{56}$Ni mixing is more noticeable, in 
our models than spherical mixing cases. 
For example, for Model A, the cut-off energy (by photoelectric absorptions) 
is already at $\sim 100$ keV at the emergence of the $\gamma$-rays, and 
it does not temporally evolve very much, because the mass fraction of 
$^{56}$Ni (Co) does not increase very much toward the center along the $z$-axis. 
This constant cut-off energy would be useful to distinguish the aspherical models 
from spherical ones, once we have a sequence of hard X-ray observations. 

Another interesting result is that line profiles are sensitively dependent 
on asphericity, as well as observer's direction. 
Lines should show larger amount of blueshift for an observer closer 
to the polar ($z$) direction. For an observer at equatorial ($r$) direction, 
lines show emission at the rest wavelength even at very early epochs. 
These are not expected in spherically symmetric models. 
Although analysis of line profiles will need future observatories 
with very high sensitivity (e.g., Takahashi et al. 2001) and/or 
a supernova very nearby, once it becomes possible then it will be extremely useful 
to trace asphericity and energy, therefore the explosion mechanism, 
of supernovae. 

For example, a possibly interesting observational strategy is implied. 
In Model A the $^{56}$Ni region becomes optically thin at a rather early phase 
when the inner regions are still optically thick. It will lead to the following 
temporal evolution for an observer at the $z$-direction. 
First, a line becomes redder and redder with time. Then 
at some epoch it stops changing the shape very much, presenting closely the 
$^{56}$Ni distribution in the hemisphere moving toward the observer. 
Finally, the line from the other hemisphere appears. 
This evolution is unique as compared with spherical models or for an observer 
at $r$-direction, either of which should show continuous reddening until 
entire ejecta become optically thin. 

Because we show that there are features of the aspherical models, some of which 
depend on the degree of asphericity but not on the observer's direction, while 
the others depend on both the asphericity and the direction, combination 
of various analyses will be helpful to distinguish models and observer's directions. 
A problem is, of course, how many hypernovae are expected to be able to be 
observed in current and future observatories. 
For the aspherical model A with $E_{51} = 20$, 
the expected maximum line fluxes are about a half of those of typical SN Ia models. 
The less energetic model ($E_{51} = 10$) yields even smaller fluxes. 
This is because of smaller amount of $^{56}$Ni and larger ratio ${M_{\rm ej}}^2/E_{\rm K}$ 
(and therefore later emergence of the radioactive decay lines) than the SN Ia models. 
Even worse, the occurring rate is much smaller for hypernovae than SNe Ia. 
In this respect, there is no doubt that SNe Ia are most promising targets 
in radioactive $\gamma$-rays. 

However, hypernovae could still be an interesting targets in $\gamma$-rays 
among core-collapse supernovae. Taking SN 1987A as a typical SN II, 
its peak 847 keV line flux was $\sim 10^{-3}$ photons cm$^{-2}$ s$^{-1}$, 
yielding $\sim 3 \times 10^{-8}$ cm$^{-2}$ s$^{-1}$ if it were at 10 Mpc.  
Our aspherical hypernova models predict the peak flux more than two orders of magnitude 
larger than this. Therefore, for a given sensitivity of instruments, 
it is visible at least an order of magnitude farther than a typical SN II. 
Low mass SNe Ib/c should also be discussed. 
Although the nature of the SNe Ib/c seems very heterogeneous, 
let us assume that $M_{\rm ej} = 2\Msun$, $E_{51} =1$, and $M$($^{56}$Ni) $= 0.07\Msun$. 
These values give the $\gamma$-ray escaping time scale comparable to 
the present models (see \S 5), resulting in the estimate of the peak line flux 
$\sim 5$ times smaller than the present models. 
For the sensitivity of the {\it INTEGRAL}, we estimate that 
the maximum distances within which $\gamma$-rays are detectable 
are $\sim 500$ kpc, $4$ Mpc, and $7$ Mpc, for 
SNe II, Ib/c, and hypernovae, respectively 
(here we assume the line width 10, 20, and 40 keV for SNe II, Ib/c, and hypernovae, 
respectively). 
The latter two, especially the detection limit for hypernovae, 
cover some star burst galaxies like M82 ($\sim 4$Mpc). 
Since SN 1987A, the nearest SNe for each type are 
SN II 2004dj ($\sim 3$ Mpc), SN Ic 1994I ($\sim 7$ Mpc), and 
a hypernova (but weaker than SN 1998bw) SN Ic 2002ap ($\sim 10$ Mpc). 
the prototypical hypernova SN 1998bw was at $\sim 36$ Mpc. 
Therefore, the sensitivity of the {\it INTEGRAL} is unfortunately not enough 
to detect these core-collapse supernovae, and we will have to wait for 
next generation hard X- and $\gamma$-ray telescopes. 
If these telescopes will be designed to archive sensitivity 
better than the {\it INTEGRAL} by two orders of magnitudes (see e.g., Takahashi 
et al. 2001), 
then the maximum distances are 5 Mpc, 40 Mpc, and 70 Mpc for SNe II, Ibc, and hypernovae, 
respectively. 
The detection limits for SNe Ib/c and hypernovae then cover some 
clusters of galaxies like Virgo ($\sim 18$Mpc), Fornax ($\sim 18$Mpc), 
and even Hydra ($\sim 40$Mpc). 
This sensitivity will lead to comprehensive study of 
$\gamma$-ray emission, and therefore of explosive nucleosynthesis, 
in core collapse supernovae. 

Observed occurring rate is $\sim 7 \times 10^{-3}$ and $\sim 1 \times 10^{-3}$ 
per year in an average galaxy for SNe II and SN Ib/c, respectively. 
The hypernova rate is rather uncertain. Podsiadlowski et al. (2004) 
gave a conservative estimate of the rate $\sim 10^{-5}$ per year. 
These values give, by multiplying the rate and the volume detectable, 
the observed probability $\sim 1$ (SNe II) : $70$ (SNe Ib/c, but not hypernovae) : 
$4$ (hypernovae) for a given detector. 
The difference between usual SNe Ib/c and hypernovae may (probably) be even smaller, 
since the estimate of SNe Ib/c flux above is probably the upper limit 
(i.e., $M_{\rm ej} = 2\Msun$ seems to be almost the lower limit) and 
that the hypernova rate is possibly the underestimate. 

In sum, as long as the detectability is concerned, 
a hypernova is worse than (usual) SNe Ib/c, 
but potentially better than SNe II. 
Another question is if any asphericity similar to that derived for the prototypical 
hypernova SN 1998bw (Maeda et al. 2006ab) exists in (usual) SNe Ib/c. 
If it does, as suggested by e.g., Wang et al. (2001), then most of the 
properties shown in this paper should also apply for hard X- and $\gamma$-ray 
emission from these SNe Ib/c (since the time scale is similar to hypernovae: see \S 5). 
One should of course take into account the fact that the expansion velocity and 
$M$($^{56}$Ni) are smaller than hypernovae, and ultimately needs direct computations 
based on a variety of explosion models.

\acknowledgements

The author is supported through the JSPS 
(Japan Society for the Promotion of Science) 
Research Fellowship for Young Scientists.

\onecolumn

\clearpage
\begin{figure}
\begin{center}
	\begin{minipage}[t]{0.4\textwidth}
		\epsscale{1.0}
		\plotone{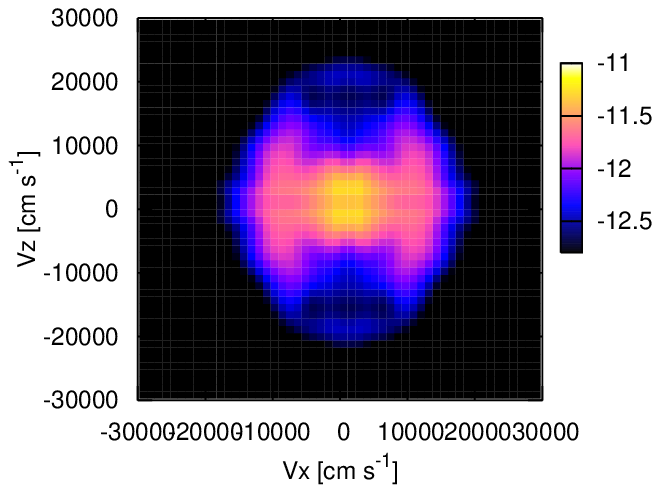}
	\end{minipage}
	\begin{minipage}[t]{0.4\textwidth}
		\epsscale{1.0}
		\plotone{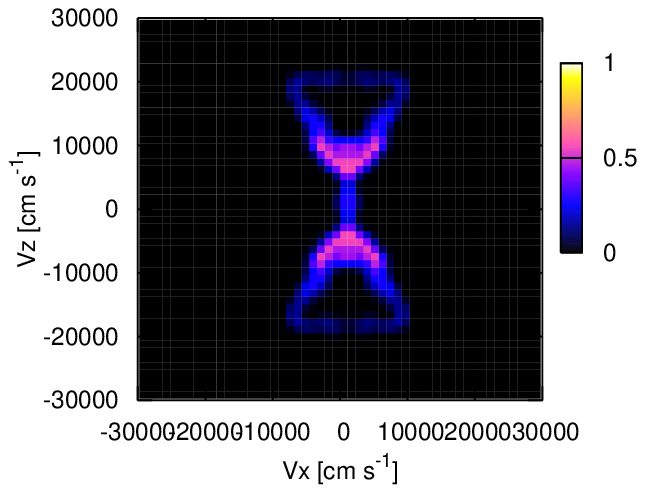}
	\end{minipage}
\end{center}
\caption[]
{Model A (Maeda et al. 2002) with $E_{51}=20$. 
The left panel shows density distribution in a logarithmic scale 
at 10 days after the explosion. The ejecta is already 
in a homologous expansion phase, so that the distribution is 
shown in the velocity space. The right panel shows mass fractions 
of $^{56}$Ni in a linear scale.  
\label{fig1}}
\end{figure}

\clearpage
\begin{figure}
\begin{center}
	\begin{minipage}[t]{0.4\textwidth}
		\epsscale{1.0}
		\plotone{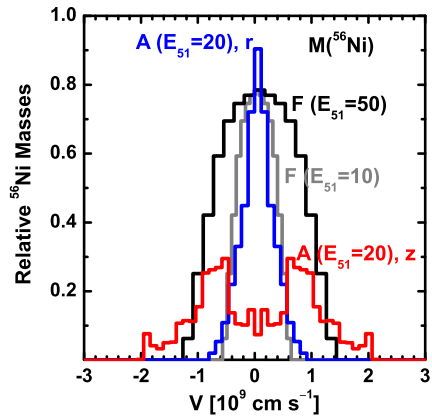}
	\end{minipage}
\end{center}
\caption[]
{Line-of-sight $^{56}$Ni distribution for Model F with $E_{51} = 10$ 
(gray), F with $E_{51} = 50$ (black), and for Model A 
with $E_{51} = 20$ along the polar ($z$) direction (red) and 
along the equator ($r$) direction (blue). 
Masses of $^{56}$Ni within the line-of-sight velocity $V \sim V+dV$ 
(with $dV$ constant) are shown as a function of the line-of-sight velocity. 
The mass is in an arbitrary unit. 
\label{fig2}}
\end{figure}

\clearpage
\begin{figure}
\begin{center}
	\begin{minipage}[t]{0.8\textwidth}
		\epsscale{1.0}
		\plotone{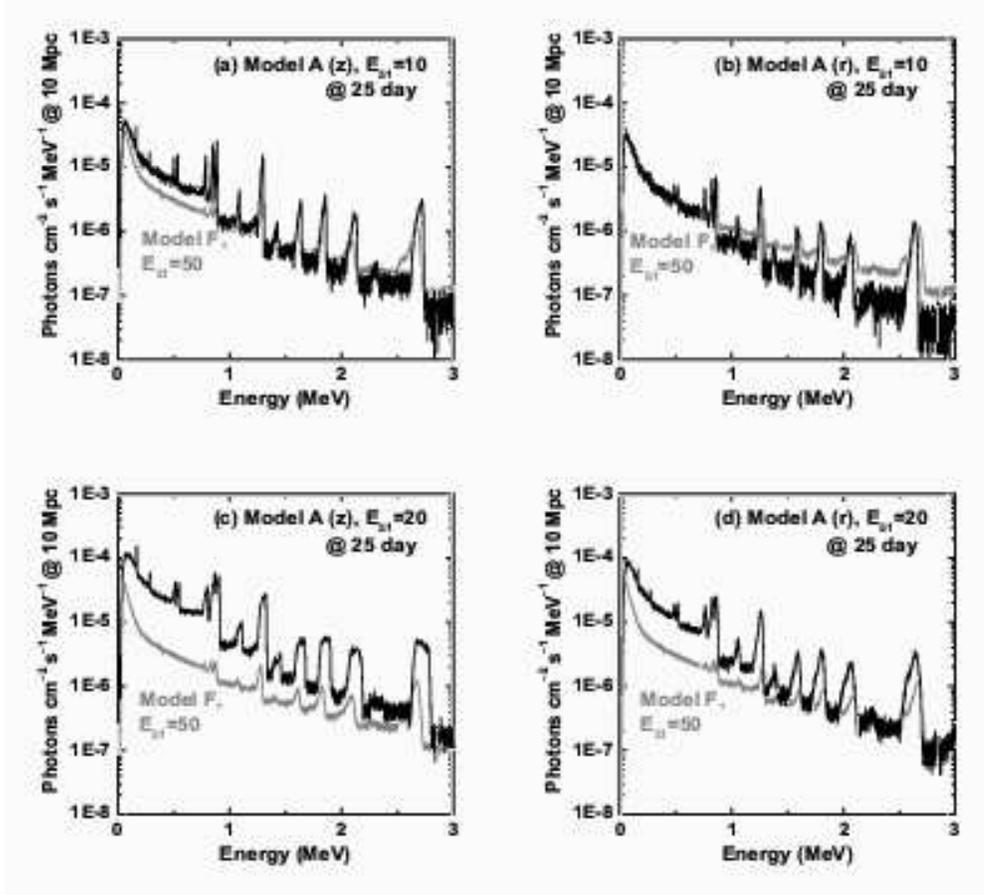}
	\end{minipage}
\end{center}
\caption[]
{Synthetic hard X-ray and $\gamma$-ray spectra at 25 days after the explosion 
at the reference distance $10$ Mpc. 
The spectra are shown for Model A ((a,b) $E_{51}$ = 10; (c,d) $E_{51} = 20$).  
The observer's direction is along the $z$-axis (a, c) and along the $r$-axis 
(b, d). In each panel, a synthetic spectrum for Model F ($E_{51} = 50$) at 25 days 
is also shown (gray) for comparison. Model F with $E_{51} = 10$ is still 
extremely optically thick in these wavelengths at this epoch, 
and therefore not visible. 
\label{fig3}}
\end{figure}

\clearpage
\begin{figure}
\begin{center}
	\begin{minipage}[t]{0.8\textwidth}
		\epsscale{1.0}
		\plotone{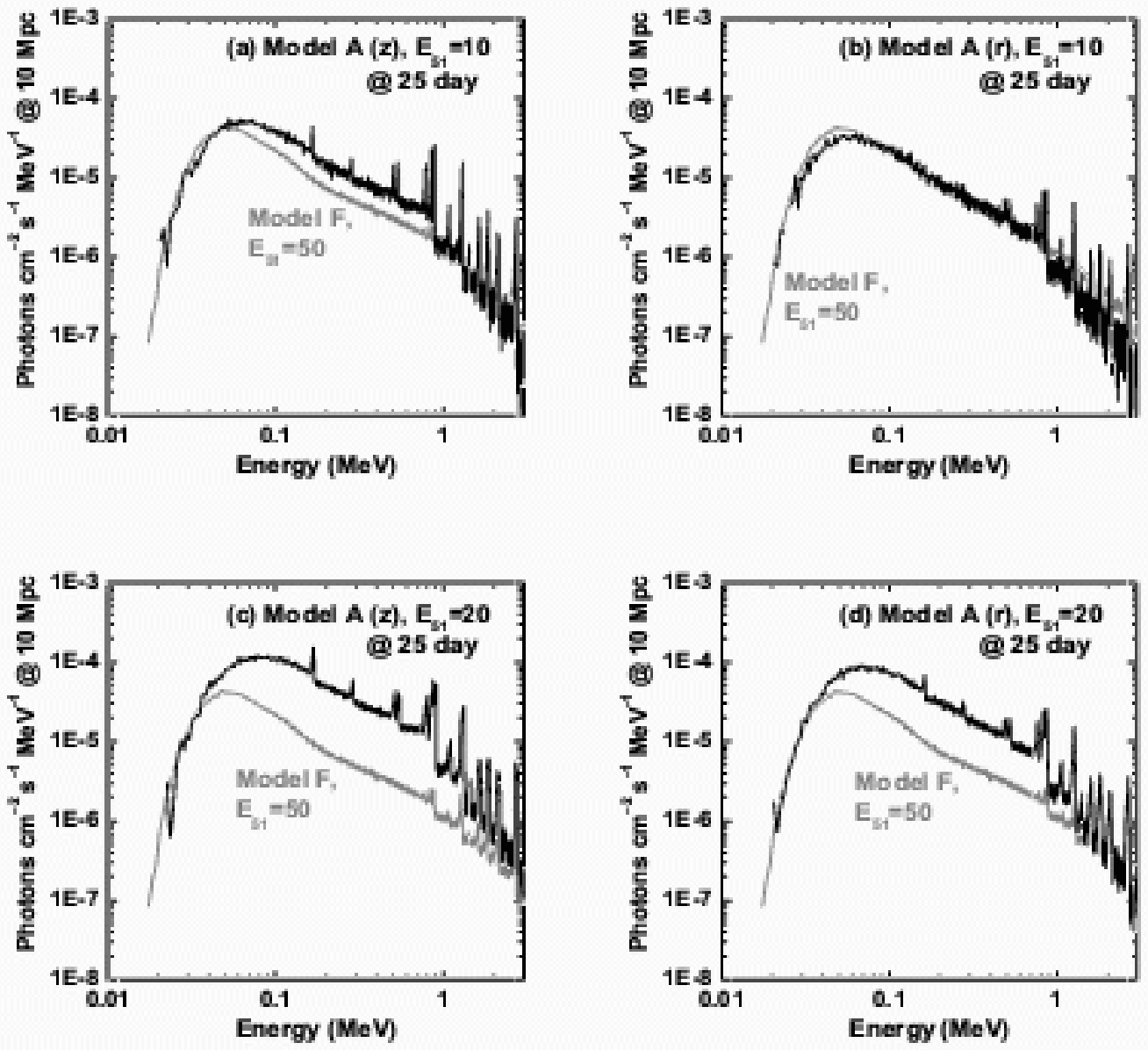}
	\end{minipage}
\end{center}
\caption[]
{Same as Figure 3 but with the horizontal axis in logarithmic scale, 
for presentation of the hard X-ray spectra. 
\label{fig4}}
\end{figure}

\clearpage
\begin{figure}
\begin{center}
	\begin{minipage}[t]{0.8\textwidth}
		\epsscale{1.0}
		\plotone{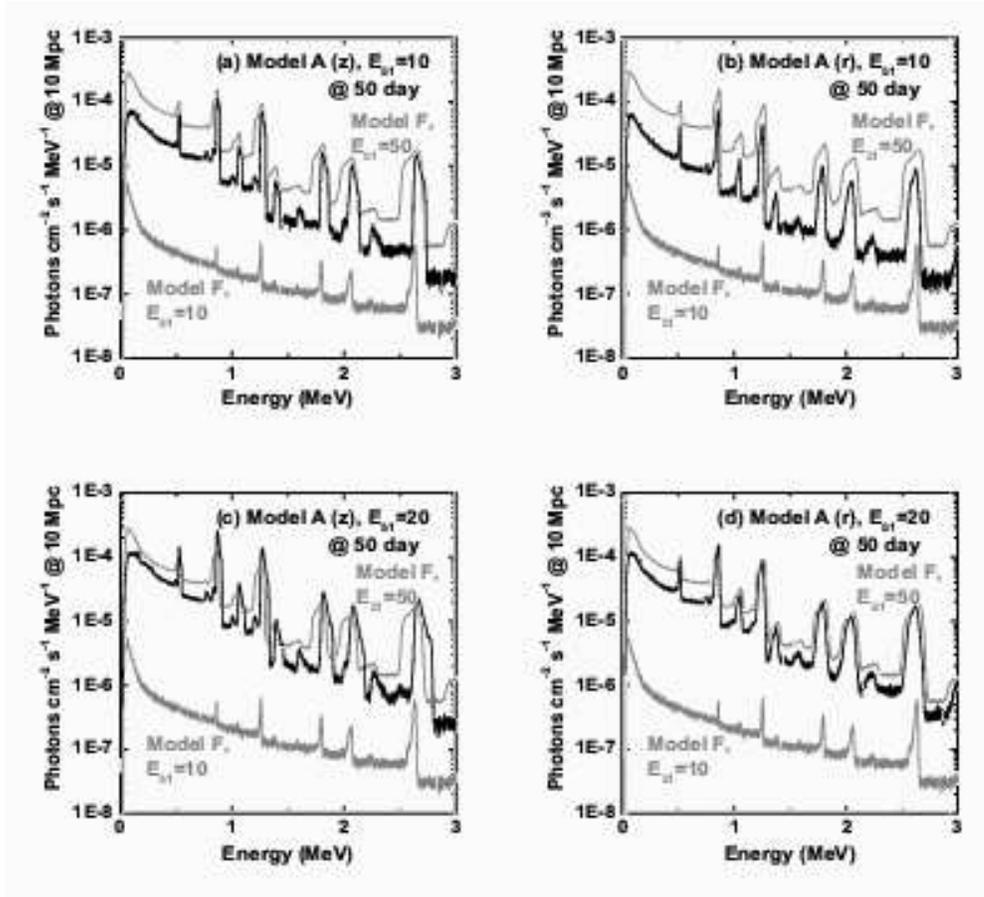}
	\end{minipage}
\end{center}
\caption[]
{Synthetic hard X-ray and $\gamma$-ray spectra at 50 days after the explosion 
at the reference distance $10$ Mpc. 
See the caption of Figure 3. Here, two spherical models (Model F with $E_{51} 
= 10$ (lower) and $50$ (upper)) are shown for comparison. 
\label{fig5}}
\end{figure}

\clearpage
\begin{figure}
\begin{center}
	\begin{minipage}[t]{0.8\textwidth}
		\epsscale{1.0}
		\plotone{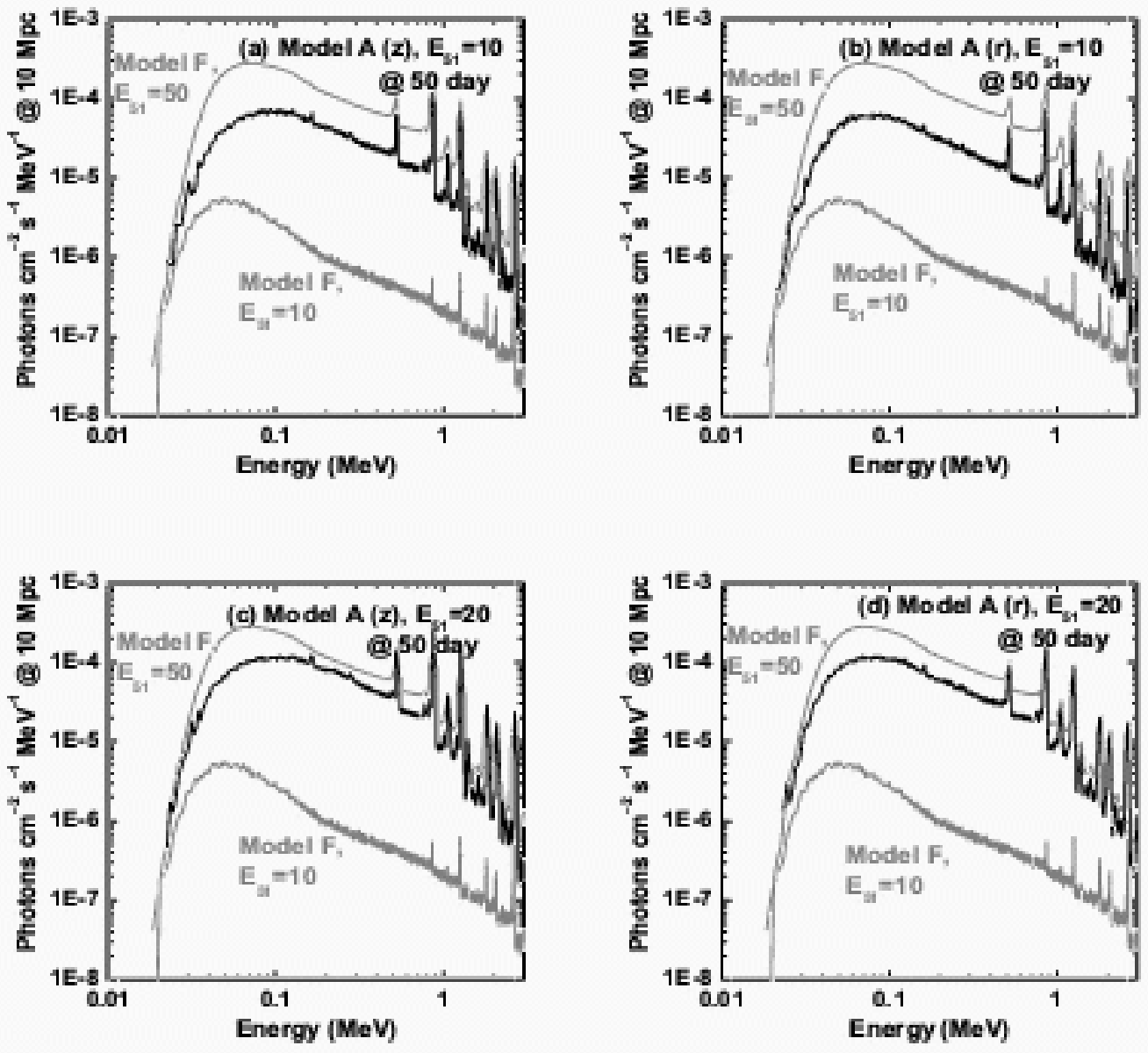}
	\end{minipage}
\end{center}
\caption[]
{Same as Figure 5 but with the horizontal axis in logarithmic scale, 
for presentation of the hard X-ray spectra. 
\label{fig6}}
\end{figure}

\clearpage
\begin{figure}
\begin{center}
	\begin{minipage}[t]{0.4\textwidth}
		\epsscale{1.0}
		\plotone{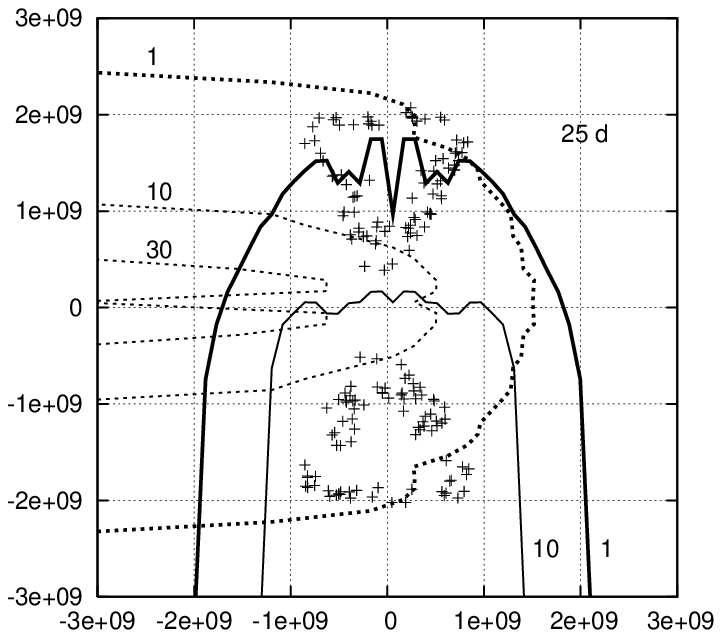}
	\end{minipage}
	\begin{minipage}[t]{0.4\textwidth}
		\epsscale{1.0}
		\plotone{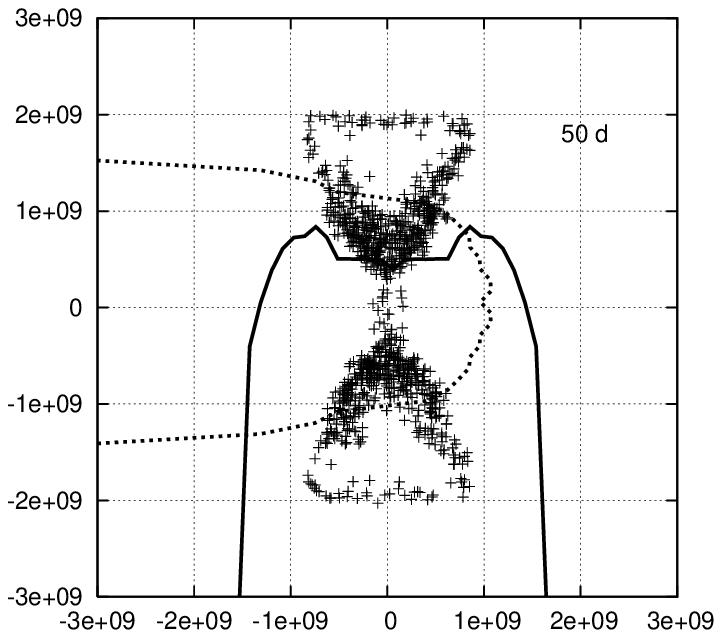}
	\end{minipage}\\
	\begin{minipage}[t]{0.4\textwidth}
		\epsscale{1.0}
		\plotone{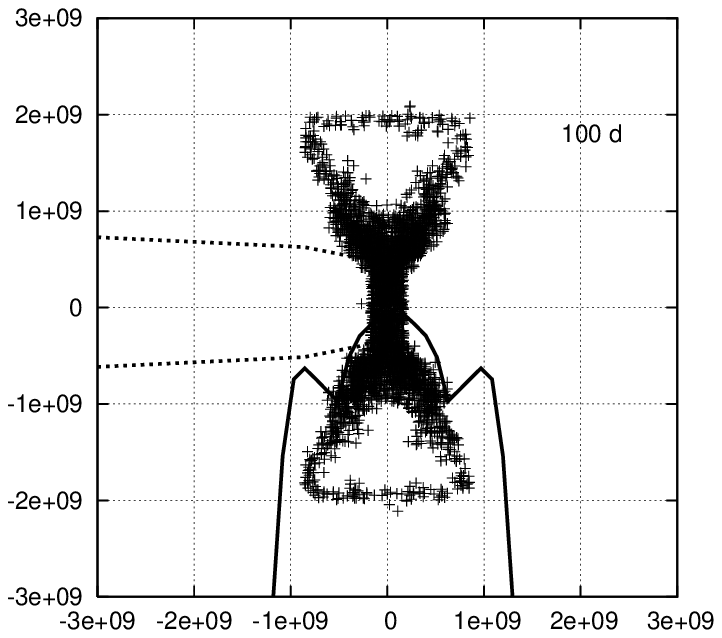}
	\end{minipage}
	\begin{minipage}[t]{0.4\textwidth}
		\epsscale{1.0}
		\plotone{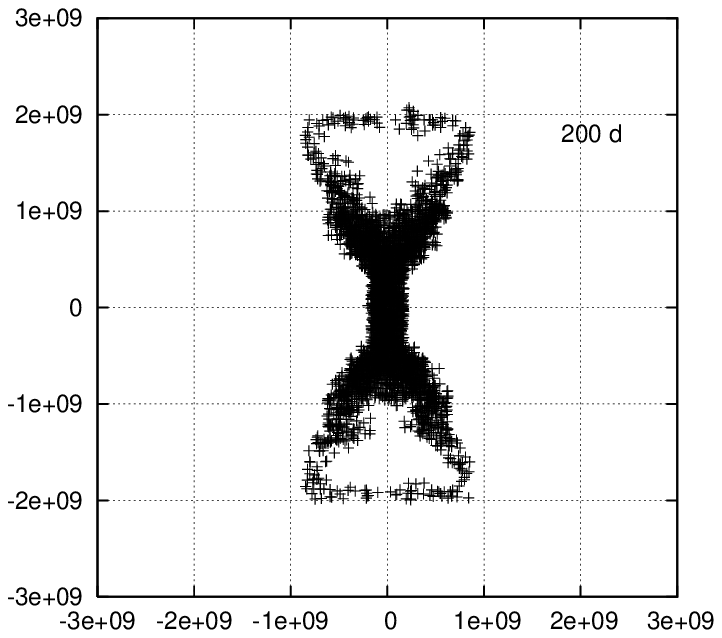}
	\end{minipage}
\end{center}
\caption[]
{Last scattering points of $\gamma$ and hard X-ray photons at 25, 50, 100, 
and 200 days since the explosion for Model A with $E_{51} = 20$. 
Also shown are contours of optical depth for an observer at $z$-axis 
(at $+z$ direction: solid) and at $r$-axis (at $+r$ direction: dotted). 
The horizontal and vertical axes are 
respectively the $V_{x}$ and the $V_{z}$ axes. 
At 25 days, the contour is shown for the optical depth $\tau_{\gamma} = $
1, 10, and 30, while for other epochs it is shown only for $\tau_{\gamma} = 1$. 
The optical depth for the contour is here computed assuming the opacity 
equal to 1/6 of the Thomson cross section. 
\label{fig7}}
\end{figure}

\clearpage
\begin{figure}
\begin{center}
	\begin{minipage}[t]{0.4\textwidth}
		\epsscale{1.0}
		\plotone{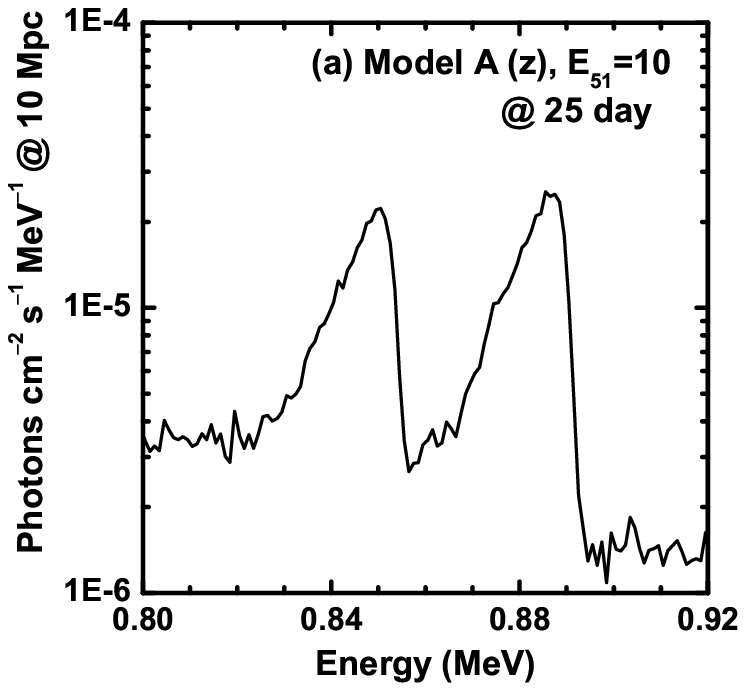}
	\end{minipage}
	\begin{minipage}[t]{0.4\textwidth}
		\epsscale{1.0}
		\plotone{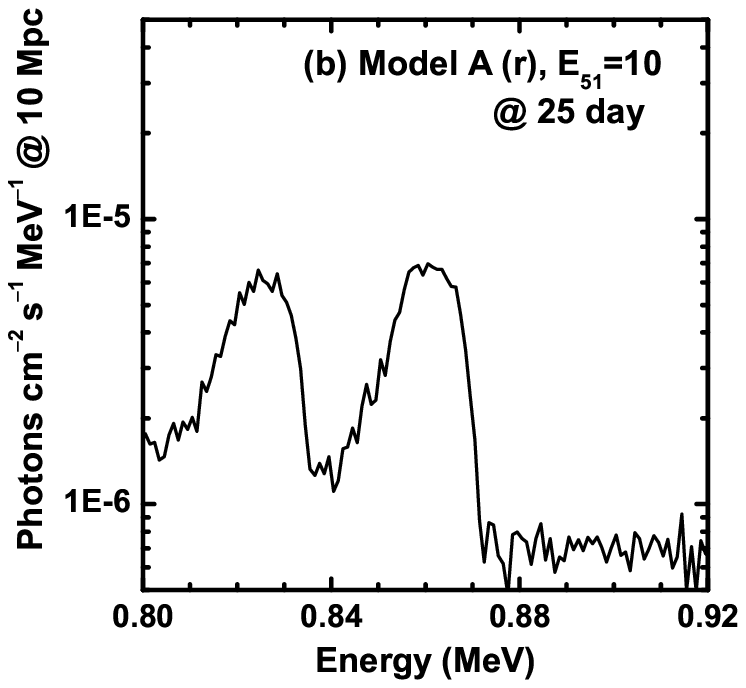}
	\end{minipage}\\
	\begin{minipage}[t]{0.4\textwidth}
		\epsscale{1.0}
		\plotone{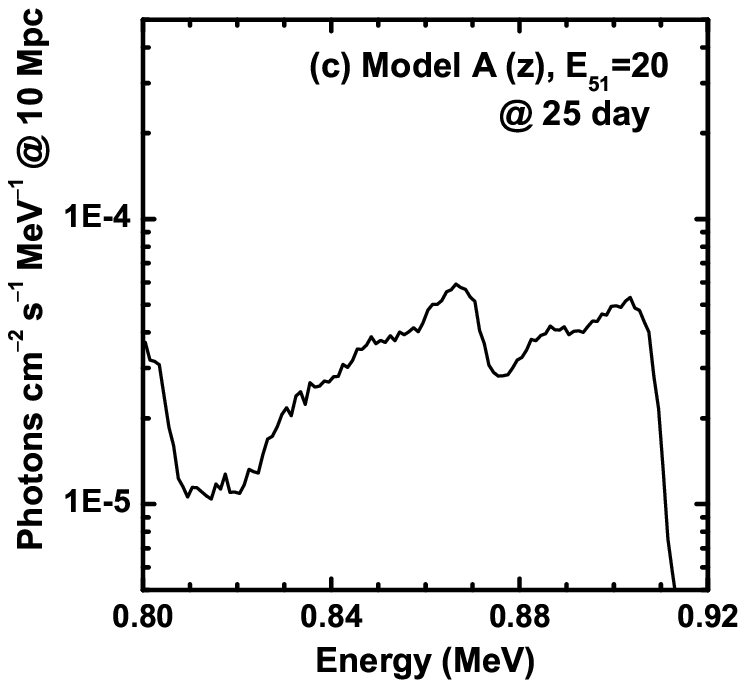}
	\end{minipage}
	\begin{minipage}[t]{0.4\textwidth}
		\epsscale{1.0}
		\plotone{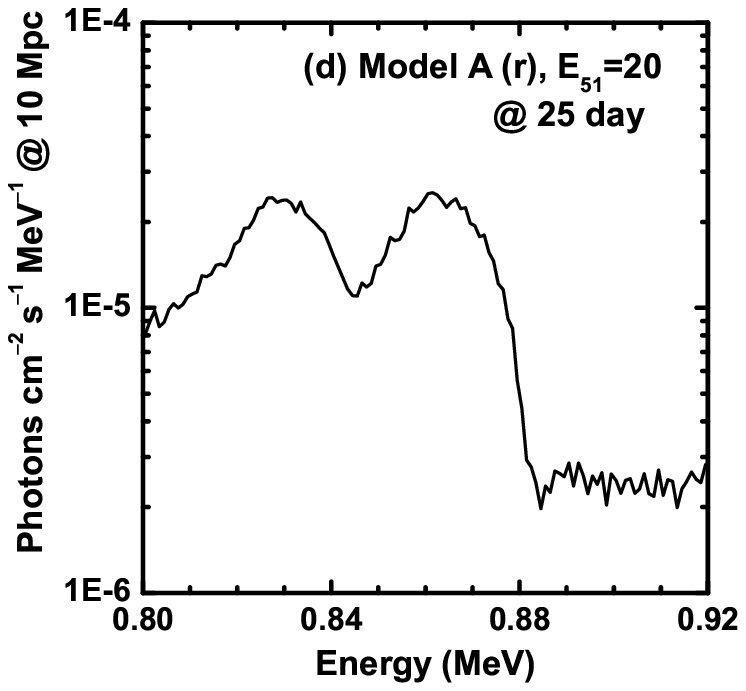}
	\end{minipage}\\
	\begin{minipage}[t]{0.4\textwidth}
		\epsscale{1.0}
		\plotone{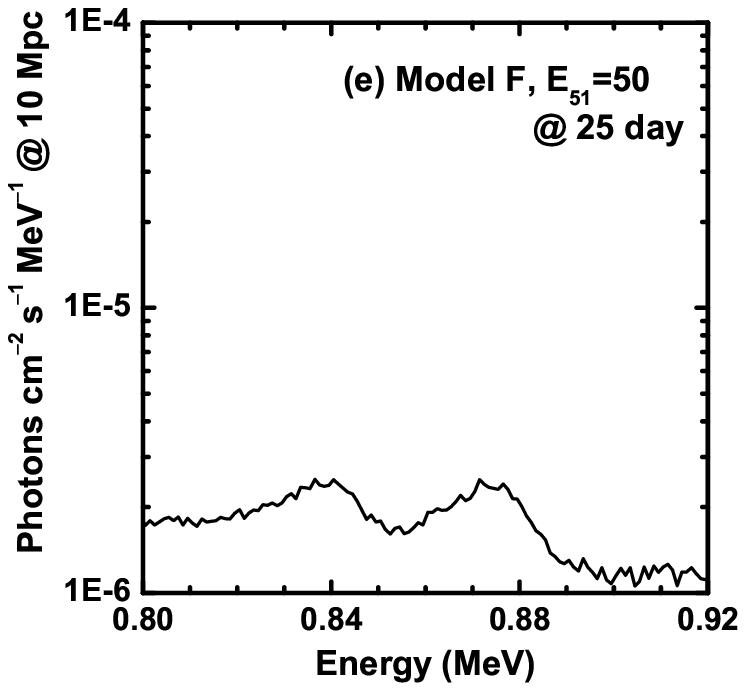}
	\end{minipage}
\end{center}
\caption[]
{Line emission of $^{56}$Ni 812 KeV and $^{56}$Co 847 KeV at 25 days 
after the explosion at the reference distance $10$ Mpc. Models spectra are shown for Model A 
with $E_{51} = 10$ (a,b) and $E_{51} = 20$ (c,d), 
with the observer at the $z$-axis (a,c) and at the $r$-axis (b,d). 
Also shown is a spectrum of Model F with $E_{51} = 50$ (e). 
\label{fig8}}
\end{figure}

\clearpage
\begin{figure}
\begin{center}
	\begin{minipage}[t]{0.4\textwidth}
		\epsscale{1.0}
		\plotone{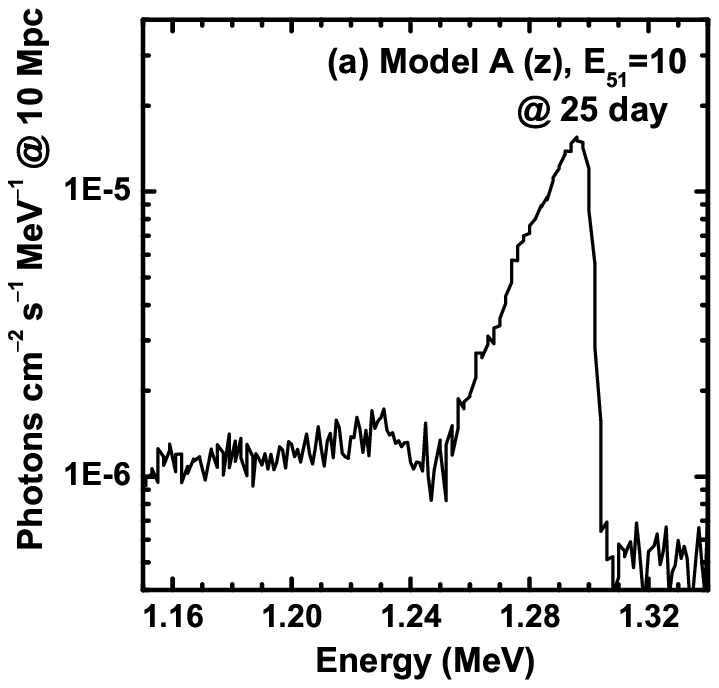}
	\end{minipage}
	\begin{minipage}[t]{0.4\textwidth}
		\epsscale{1.0}
		\plotone{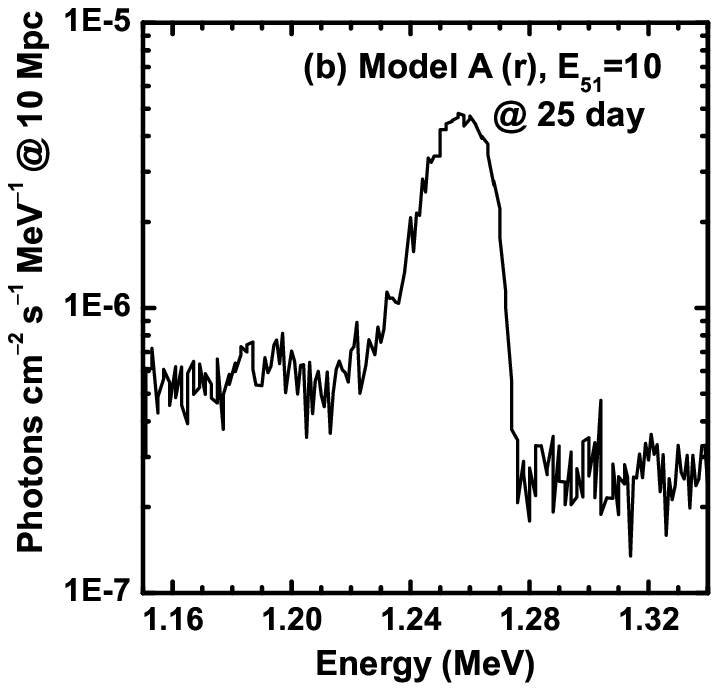}
	\end{minipage}\\
	\begin{minipage}[t]{0.4\textwidth}
		\epsscale{1.0}
		\plotone{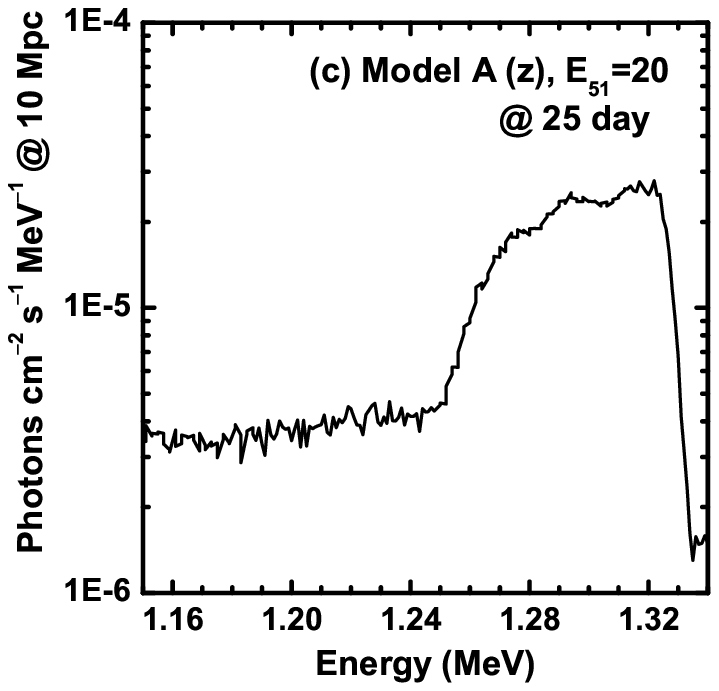}
	\end{minipage}
	\begin{minipage}[t]{0.4\textwidth}
		\epsscale{1.0}
		\plotone{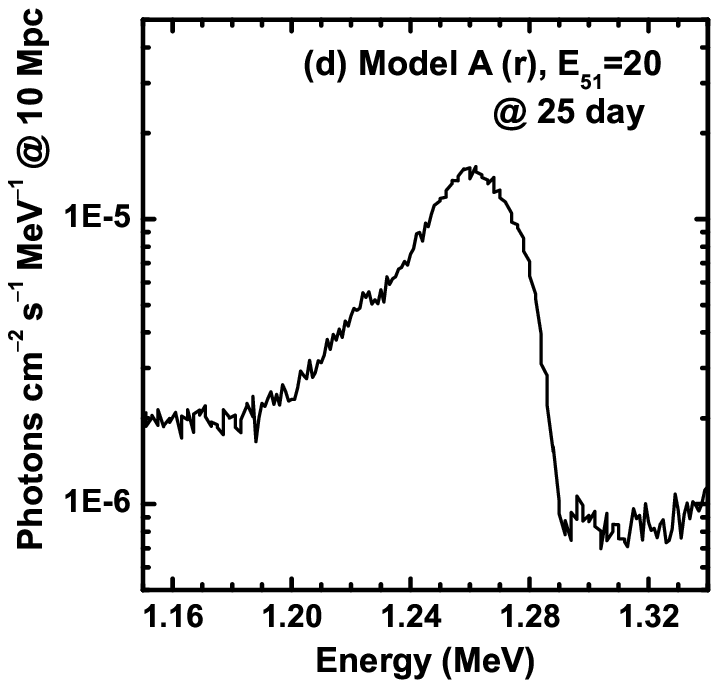}
	\end{minipage}\\
	\begin{minipage}[t]{0.4\textwidth}
		\epsscale{1.0}
		\plotone{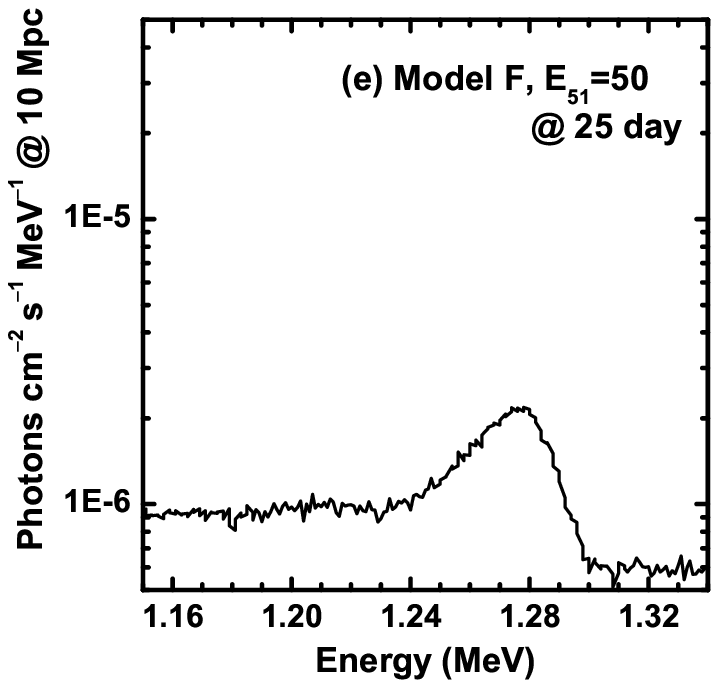}
	\end{minipage}
\end{center}
\caption[]
{Line emission of $^{56}$Co 1238 KeV at 25 days 
after the explosion at the reference distance $10$ Mpc. See the caption of Figure 8. 
\label{fig9}}
\end{figure}

\clearpage
\begin{figure}
\begin{center}
	\begin{minipage}[t]{0.4\textwidth}
		\epsscale{1.0}
		\plotone{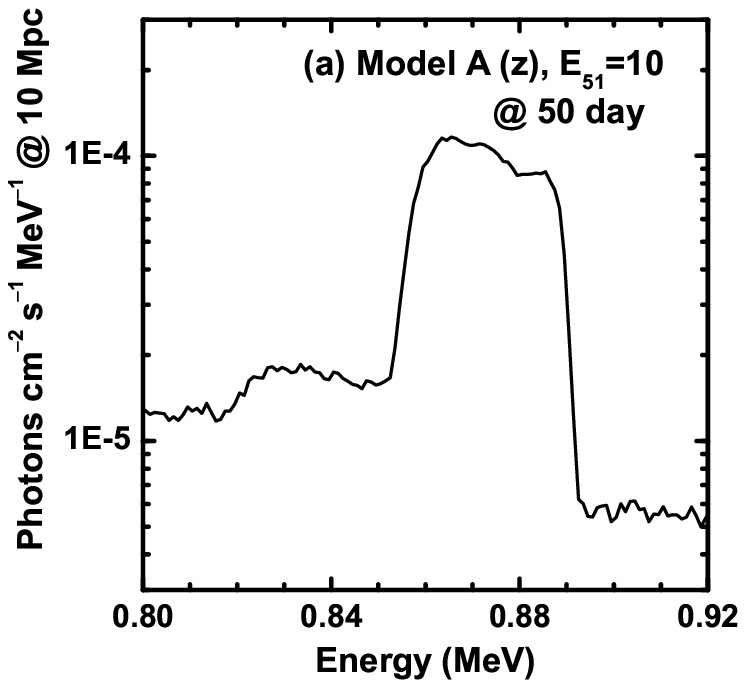}
	\end{minipage}
	\begin{minipage}[t]{0.4\textwidth}
		\epsscale{1.0}
		\plotone{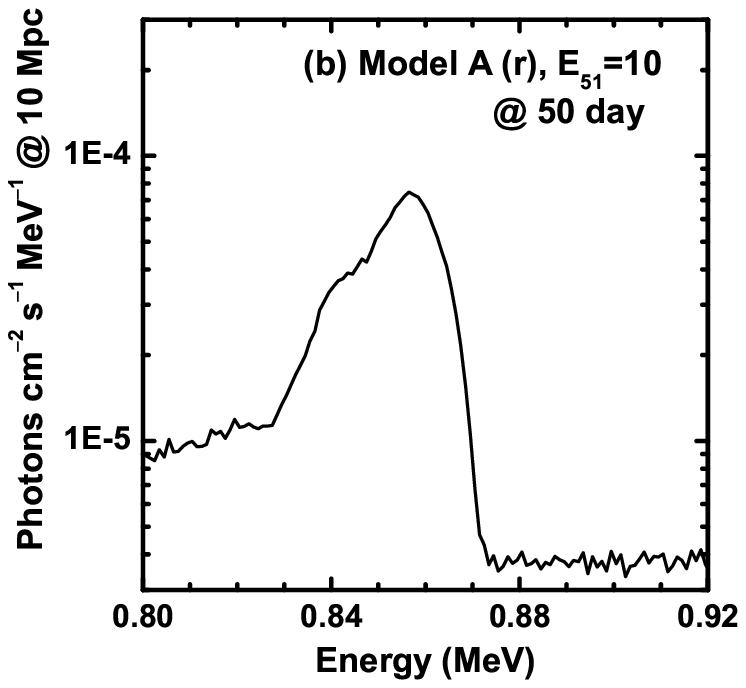}
	\end{minipage}\\
	\begin{minipage}[t]{0.4\textwidth}
		\epsscale{1.0}
		\plotone{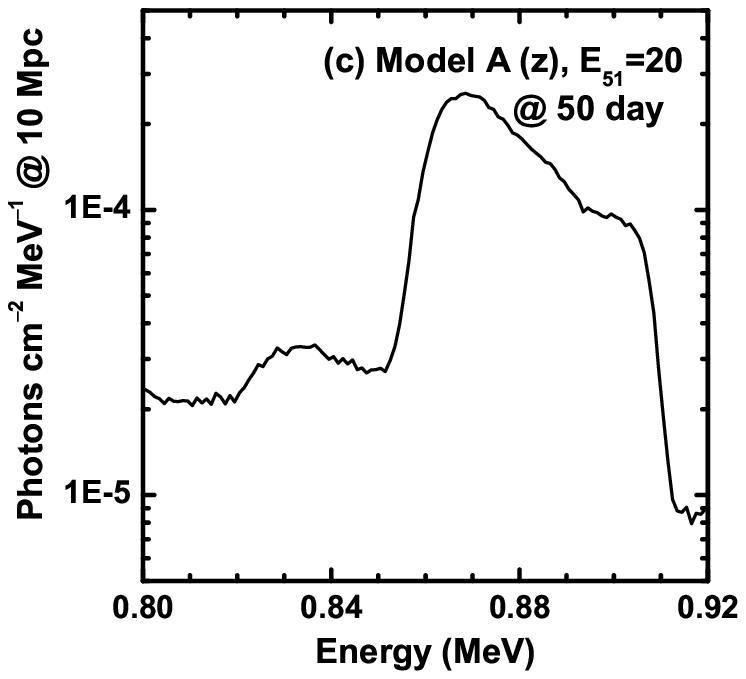}
	\end{minipage}
	\begin{minipage}[t]{0.4\textwidth}
		\epsscale{1.0}
		\plotone{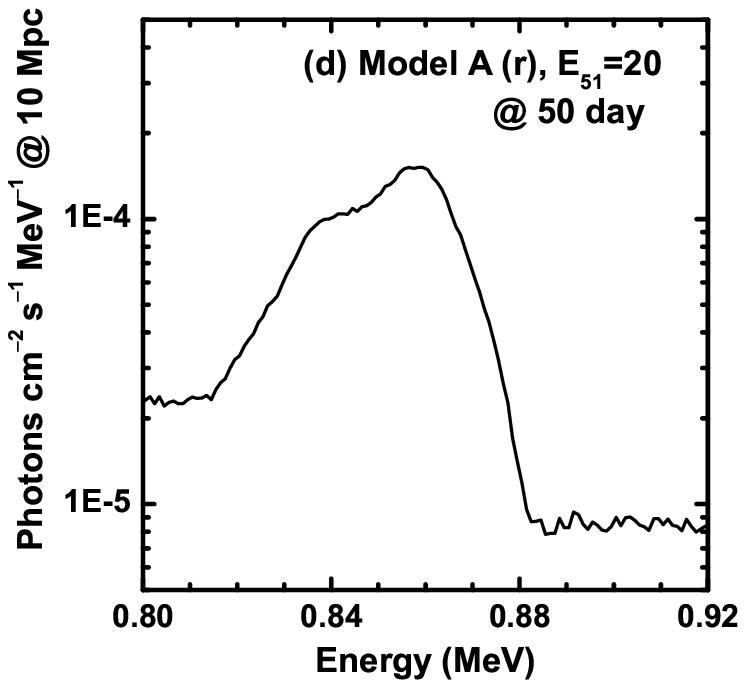}
	\end{minipage}\\
	\begin{minipage}[t]{0.4\textwidth}
		\epsscale{1.0}
		\plotone{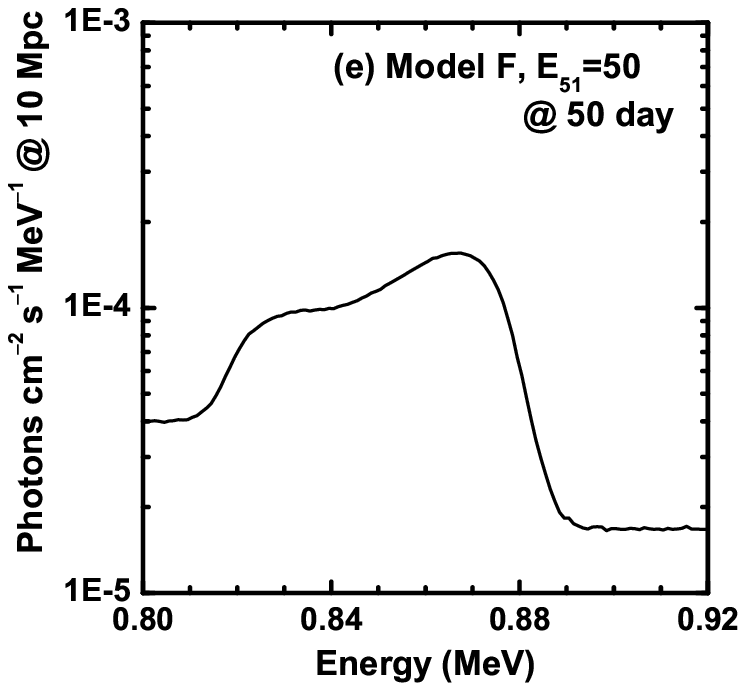}
	\end{minipage}
	\begin{minipage}[t]{0.4\textwidth}
		\epsscale{1.0}
		\plotone{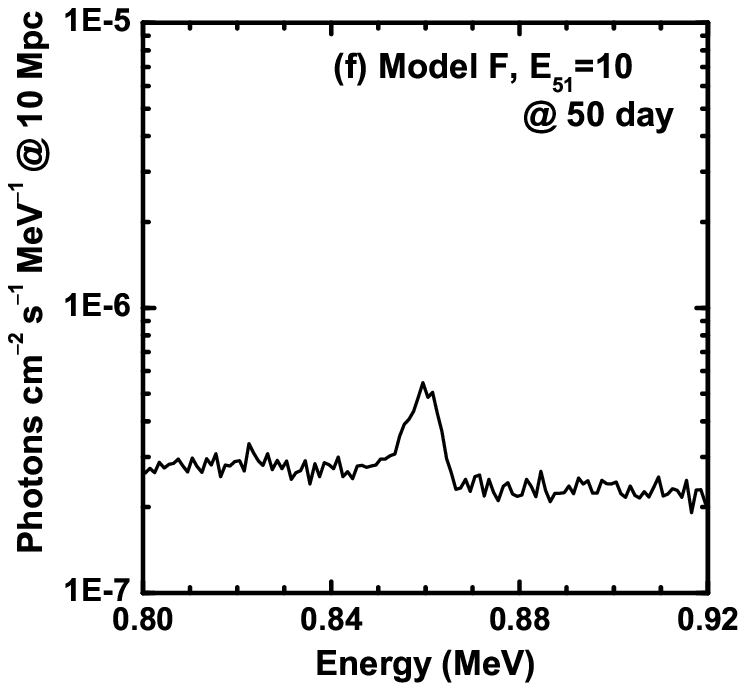}
	\end{minipage}
\end{center}
\caption[]
{Line emission of $^{56}$Ni 812 KeV and $^{56}$Co 847 KeV at 50 days 
after the explosion at the reference distance $10$ Mpc. Models spectra are shown for Model A 
with $E_{51} = 10$ (a,b) and $E_{51} = 20$ (c,d), 
with the observer at the $z$-axis (a,c) and at the $r$-axis (b,d). 
Also shown is a spectrum of Model F with $E_{51} = 50$ (e) and 
$E_{51} = 10$ (f). 
\label{fig10}}
\end{figure}

\clearpage
\begin{figure}
\begin{center}
	\begin{minipage}[t]{0.4\textwidth}
		\epsscale{1.0}
		\plotone{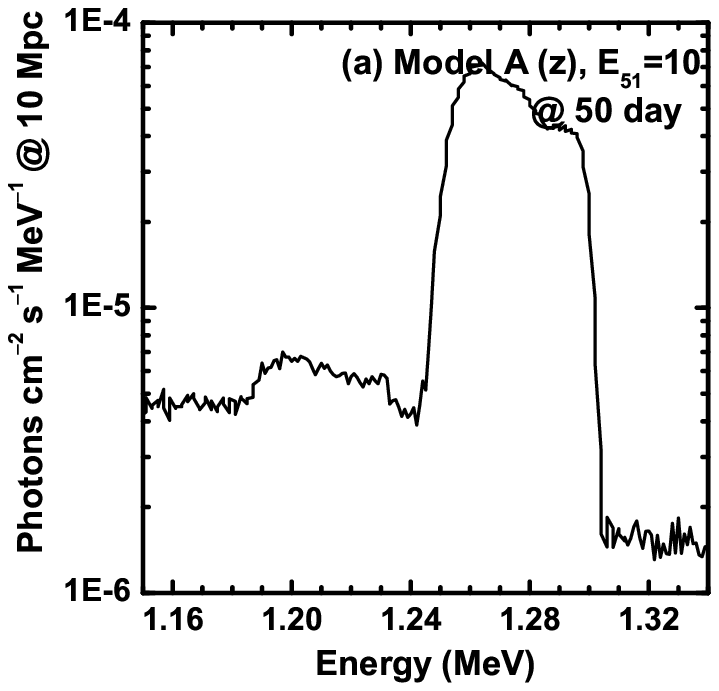}
	\end{minipage}
	\begin{minipage}[t]{0.4\textwidth}
		\epsscale{1.0}
		\plotone{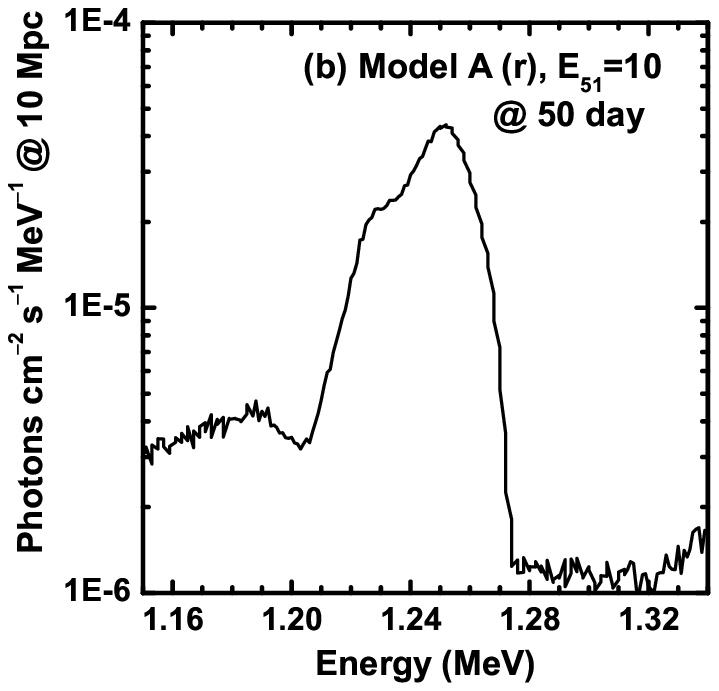}
	\end{minipage}\\
	\begin{minipage}[t]{0.4\textwidth}
		\epsscale{1.0}
		\plotone{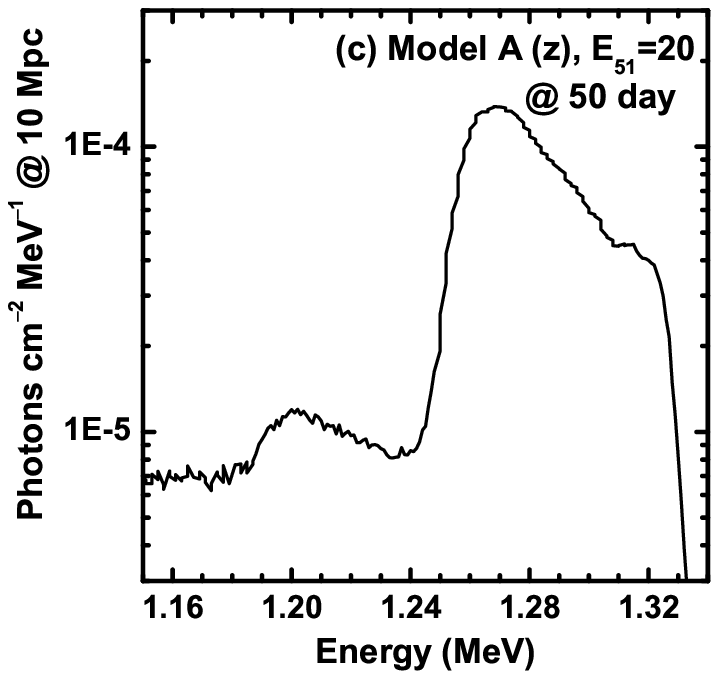}
	\end{minipage}
	\begin{minipage}[t]{0.4\textwidth}
		\epsscale{1.0}
		\plotone{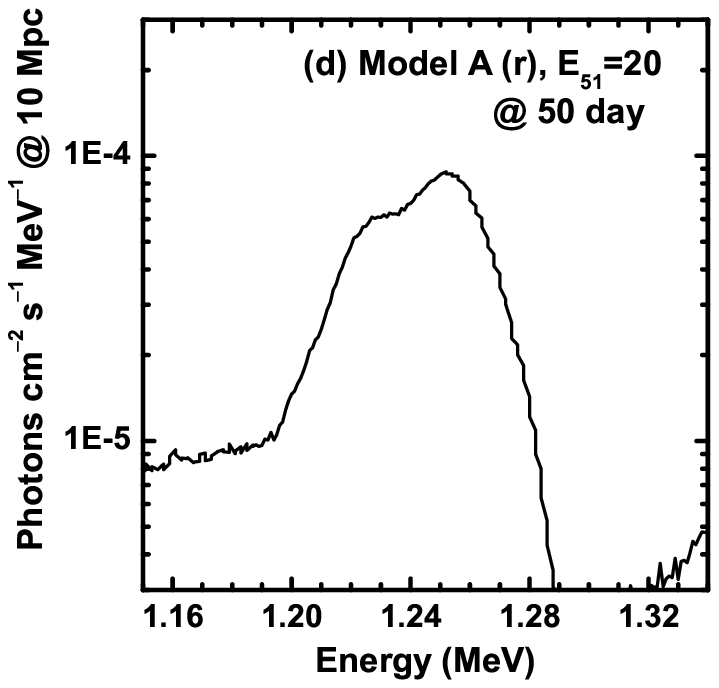}
	\end{minipage}\\
	\begin{minipage}[t]{0.4\textwidth}
		\epsscale{1.0}
		\plotone{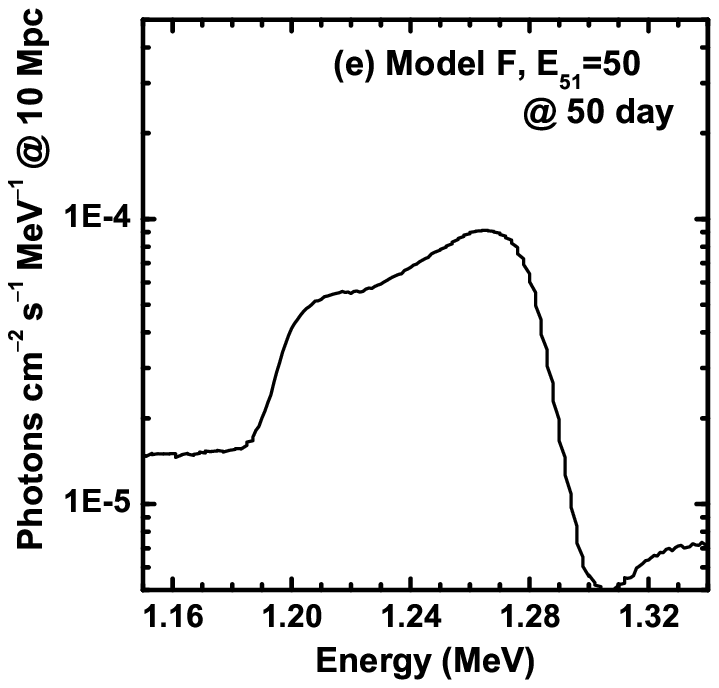}
	\end{minipage}
	\begin{minipage}[t]{0.4\textwidth}
		\epsscale{1.0}
		\plotone{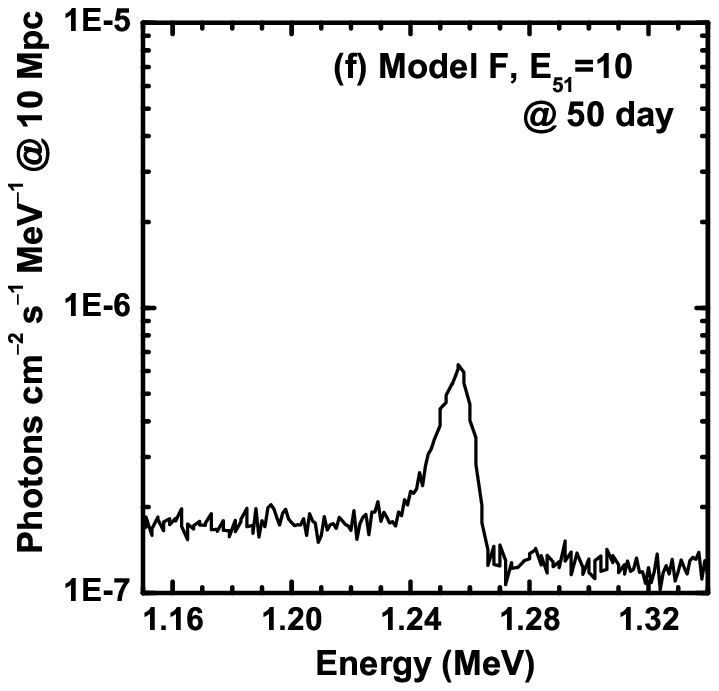}
	\end{minipage}
\end{center}
\caption[]
{Line emission of $^{56}$Co 1238 KeV at 50 days 
after the explosion at the reference distance $10$ Mpc. See the caption of Figure 10. 
\label{fig11}}
\end{figure}

\clearpage
\begin{figure}
\begin{center}
	\begin{minipage}[t]{0.4\textwidth}
		\epsscale{1.0}
		\plotone{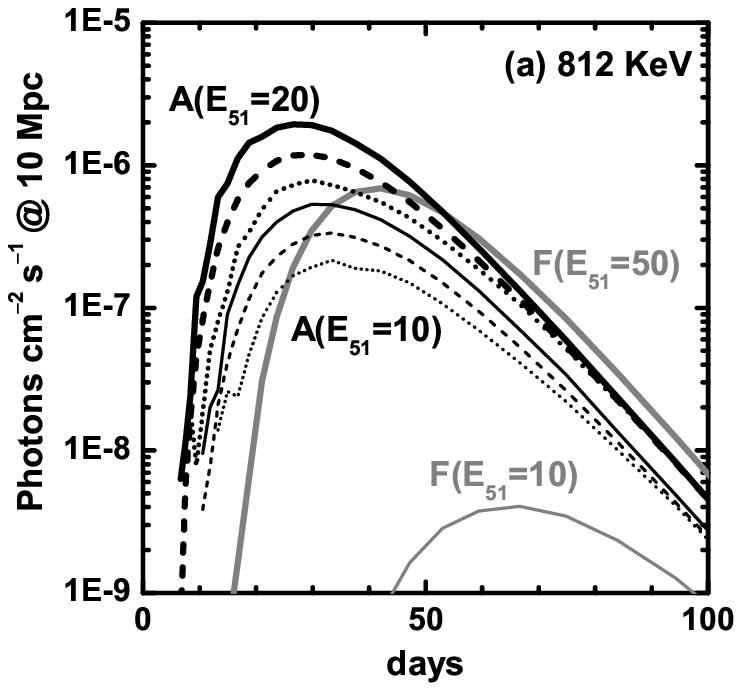}
	\end{minipage}\\
	\begin{minipage}[t]{0.4\textwidth}
		\epsscale{1.0}
		\plotone{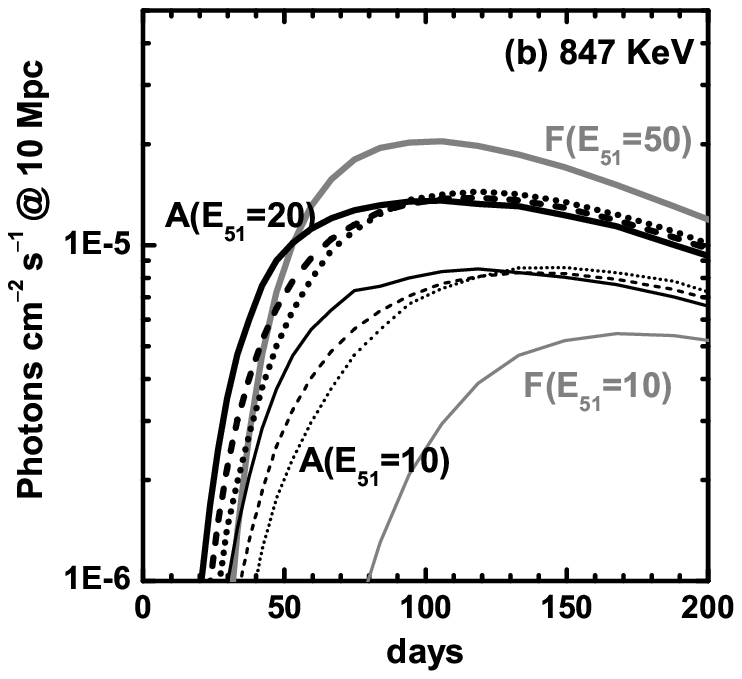}
	\end{minipage}\\
	\begin{minipage}[t]{0.4\textwidth}
		\epsscale{1.0}
		\plotone{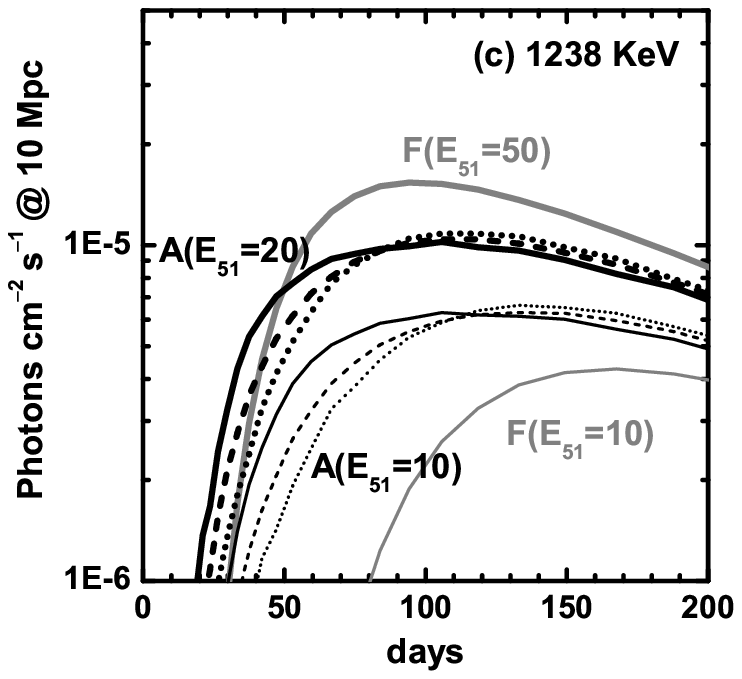}
	\end{minipage}
\end{center}
\caption[]
{Light curves of (a) the $^{56}$Ni 812 KeV line, (b) the 
$^{56}$Co 847 KeV line, and (c) the $^{56}$Co 1238 
KeV line at the reference distance $10$ Mpc. 
For the 812 \& 847 KeV lines, 
the flux is computed by integrating a spectrum in the 
$800 - 920$ KeV energy range, then by assuming 
that the escape fractions for these two lines 
are equal. For the 1238 line, the flux is computed 
by integrating a spectrum in $1150 - 1340$ KeV energy range. 
Shown here are Model A with $E_{51} = 20$ (thick-black) and $10$ 
(thin-black), and Model F with $E_{51} = 50$ (thick-gray) and 
$10$ (thin-gray). For Model A, the flux is shown for 
the $z$-direction (solid), the $r$-direction (dotted), 
as well as the angle-averaged one (dashed). 
\label{fig12}}
\end{figure}

\clearpage
\begin{figure}
\begin{center}
	\begin{minipage}[t]{0.4\textwidth}
		\epsscale{1.0}
		\plotone{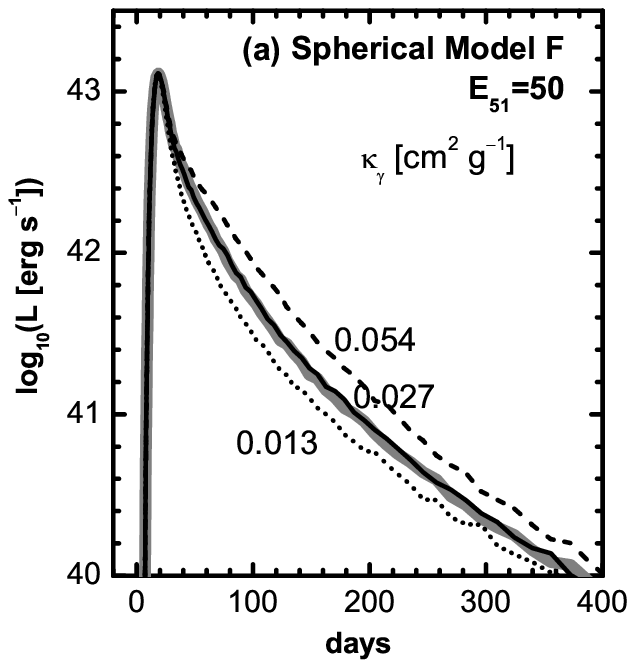}
	\end{minipage}
	\begin{minipage}[t]{0.4\textwidth}
		\epsscale{1.0}
		\plotone{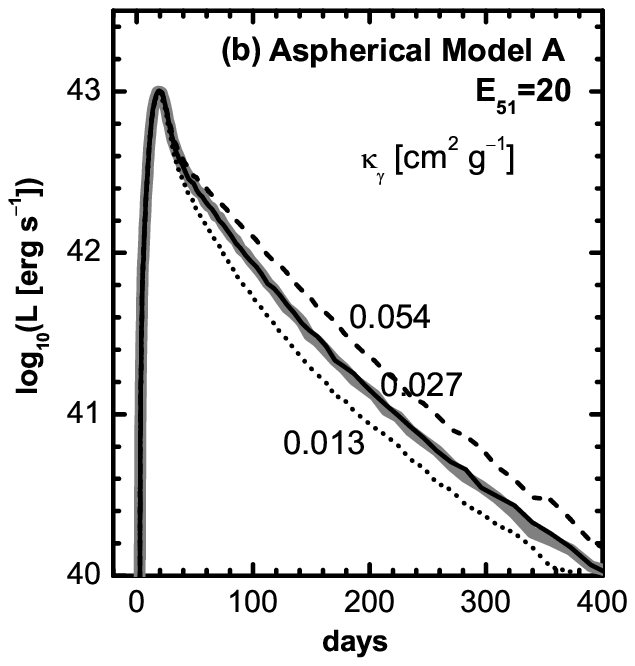}
	\end{minipage}
\end{center}
\caption[]
{Optical light curves computed for (a) Model F with $E_{51} = 50$ 
and (b) Model A with $E_{51} = 20$ 
(averaged over all the solid angles). Optical photon transport 
is solved with a simplified gray representation for 
optical opacity (Chugai 2000) by the Monte Carlo radiation transfer method 
(Maeda et al. 2006b). 
For the gamma ray transport, the detailed transport (gray) is 
compared with the simplified gray atmosphere approximation with 
the effective absorptive gamma ray opacity $\kappa_{\gamma} = 0.13$, 
$0.27$, and $0.54$ cm$^{2}$ g$^{-1}$ (black curves). 
\label{fig13}}
\end{figure}

\end{document}